\documentclass[%
reprint,
superscriptaddress,
 amsmath,amssymb,
 aps,
pra,
]{revtex4-2}

\usepackage{graphicx}% Include figure files
\usepackage{subfig}
\usepackage{dcolumn}% Align table columns on decimal point
\usepackage{bm}% bold math
\usepackage{xcolor}
\colorlet{nblue}{blue!65!cyan}
\colorlet{nred}{magenta!40!red}
\colorlet{ngreen}{green!50!nblue}
\usepackage[colorlinks,linkcolor=ngreen,citecolor=blue,urlcolor=ngreen]{hyperref}

\newcommand{\beq}{\begin{equation}}
\newcommand{\eeq}{\end{equation}}

\newcommand{\vect}[1]{\mathbf{#1}}
\newcommand{\Eq}[1]{Eq.~\eqref{#1}}

\DeclareMathOperator{\sign}{sgn}

\DeclareMathOperator{\im}{Im}
\DeclareMathOperator{\re}{Re}

\newcommand{\sgn}{\mathrm{sgn}}

\newcommand{\T}{\mathcal{T}}
\renewcommand{\S}{\mathcal{S}}
\renewcommand{\O}{\mathcal{O}}

\newcommand{\nn}{\nonumber}

\begin{document}
\title{The three-body scattering hypervolume of identical fermions in one dimension}
\author{Zipeng Wang}
\author{Shina Tan}%
 \email{shinatan@pku.edu.cn}
\affiliation{%
International Center for Quantum Materials, Peking University, Beijing 100871, China 
}%
%\collaboration{CLEO Collaboration}%\noaffiliation
\date{\today}

\begin{abstract}
We study the zero-energy collision of three identical spin-polarized fermions with short-range interactions in one dimension.
We derive the asymptotic expansions of the three-body wave function when the three fermions are far apart or one pair and the third fermion are far apart, and the three-body scattering hypervolume $D_F$ appears in the coefficients of such expansions. If the two-body interaction is attractive and supports two-body bound states, $D_F$ acquires a negative imaginary part
related to the amplitudes of the outgoing waves describing the departure of the resultant bound pair and the remaining free fermion.
For weak interaction potentials, we derive an approximate formula of the hypervolume by using the Born expansion. For the square-barrier and the square-well potentials and the Gaussian potential, we solve the three-body Schr\"{o}dinger equation to compute $D_F$ numerically.
We also calculate the shifts of energy and of pressure of spin-polarized one-dimensional Fermi gases due to a nonzero $D_F$ and the three-body recombination rate in one dimension.
\end{abstract}

\maketitle

\section{Introduction}\label{sec:intro}
One-dimensional (1D) quantum gases can be experimentally realized by applying strong confinement in two transverse directions and allow free motion along the longitudinal direction \cite{Gorlitz2001,Schreck2001,Dettmer2001,Greiner2001,Moritz2003,Paredes2004,ObservationTGgas,Syassen2008,Haller2009,PhysRevLett.104.153203}.
1D quantum gases are very different from  the ordinary three-dimensional (3D) quantum gases \cite{RevModPhys.83.1405,RevModPhys.85.1633}.
%\cite{Tonks1963,Girardeau1960,LiebLiniger1963,McGuire1964,Yang1967,Yang1969,Fermion-Boson-duality,PhysRevA.97.061603,PhysRevA.97.061605,OHYA2021168657,PhysRevLett.81.938}, 
%e.g. the Tonks–Girardeau Bose gas \cite{Tonks1963,Girardeau1960}, the Bose-Fermi duality \cite{Girardeau1960,Fermion-Boson-duality}, exactly solvable models for 1D quantum gase \cite{LiebLiniger1963,Lieb1963,McGuire1964,McGuire1965fermion1,McGuire1966fermion2,GAUDIN196755,Yang1967,Yang1969}, 1D Hubbard model \cite{hubbard,Takahashi1972,Takahashi1974}, etc.
%

The three-body problem in 1D has been studied for many years \cite{PhysRevA.72.032728,Esry2007,PhysRevA.97.061603,PhysRevA.97.061604,PhysRevA.97.061605,Quantumdroplet2018,PhysRevA.100.013614,PhysRevA.103.043307}.
In this paper, we define and study the \emph{three-body scattering hypervolume} of identical spin-polarized fermions in 1D. The scattering hypervolume is a three-body analog of the two-body scattering length \cite{tan2008three}, which can be extracted from the wave function of two particles colliding at zero energy.
If the interaction is short ranged, ie the interaction potential vanishes beyond a finite pairwise distance $r_e$,
the wave function of two particles colliding at zero energy in 1D is
\begin{equation}
\phi_l(s)=(|s|-a_{l})Y_l(s)
\end{equation}
at $|s|>r_e$ in the center-of-mass frame,
where $a_{l}$ is the  two-body scattering length in 1D, $s$ is the difference of the coordinates of the two particles, %$r_e$ is the range of interaction,
and $l$ can be 0 or 1 for $s$-wave collisions or $p$-wave collisions respectively.
$Y_0(s)=1$, and $Y_1(s)=\sign(s)$.
Here $\sgn(s)$ is the sign function. $\sgn(s)=1$ for $s>0$, $\sgn(s)=0$ for $s=0$, and $\sgn(s)=-1$ for $s<0$. 

For particles in higher dimensional spaces, people have defined and studied the three-body scattering hypervolume in various systems 
\cite{tan2008three,zhu2017threebody,mestrom2019scattering,mestrom2020van,wang2021threebody,mestrom2021pwave,mestrom2021spin1,wang2021fermion3D,wang2022fermion2D}. The three-body scattering hypervolumes have been defined and studied
for identical bosons in 3D \cite{tan2008three,zhu2017threebody,mestrom2019scattering,mestrom2020van,mestrom2021spin1}, distinguishable particles in 3D \cite{wang2021threebody,mestrom2021pwave}, identical spin-polarized fermions in 3D \cite{wang2021fermion3D} or in 2D \cite{wang2022fermion2D}.
In this paper, we define the scattering hypervolume $D_F$ of identical spin-polarized fermions in 1D, by studying the wave function of three such fermions colliding at zero energy, and study its analytical and numerical calculations and its physical implications.
Our results may be applicable to ultracold atomic Fermi gases confined in one dimension.

This paper is organized as follows.
In Sec.~\ref{sec:two-body} we define the two-body $p$-wave special functions.
In Sec.~\ref{sec:asymp} we derive the asymptotic expansions of the three-body wave function for zero energy collision. The scattering hypervolume $D_F$ appears in the coefficients in these expansions. %If the two-body interaction potential is sufficiently attractive such that it supports at least one two-body bound state, we show that $D_F$ acquires a negative imaginary part related to the amplitudes of
%the outgoing waves describing the departure of the resultant bound pair and the remaining free fermion.
In Sec.~\ref{sec:value-DF}, we derive an approximate formula of $D_F$ for weak interaction potentials by using the Born expansion. For the square-barrier and the square-well potentials and the Gaussian potential we numerically compute $D_F$ for various interaction strengths.
In Sec.~\ref{energy} we consider the dilute spin-polarized Fermi gas in 1D and derive the shifts of its energy and pressure due to a nonzero $D_F$.
In Sec.~\ref{sec:recombination}, we derive the formula for the three-body recombination rate of the dilute spin-polarized Fermi gas in 1D in terms of the imaginary part of $D_F$.

\section{two-body special functions}\label{sec:two-body}
For identical spin-polarized fermions in 1D, the $s$-wave two-body scattering is forbidden due to Fermi statistics, and only the $p$-wave scattering is permitted. 
The two-fermion scattering wave function $\Phi$ in the center-of-mass frame with collision energy $E=\hbar^2k^2/m$,
where $m$ is the mass of each fermion and $\hbar$ is Planck's constant over $2\pi$,
satisfies the following Schr\"{o}dinger equation:
\begin{equation}\label{two-body-equ}
\frac{d^2\Phi(s)}{ds^2}+\left[k^2-\frac{mV(s)}{\hbar^2}\right]\Phi(s)=0,
\end{equation}
where $V(s)$ is the two-body interaction potential.
%where $s$ is the difference of the coordinates of the two particles.
We assume that $V(s)$ is an even function of $s$, namely $V(s)=V(|s|)$, and that it vanishes at $|s|>r_e$.
At $|s|>r_e$,
\Eq{two-body-equ} is simplified as 
$\frac{d^2\Phi}{ds^2}+k^2\Phi=0$, and its solution is 
\begin{equation}
	\Phi(s)=A \sin\left(k|s|+\delta_p\right)\sign(s),\label{two-body-Phi}
\end{equation}
where %, and we define 
%$Y_0(s)\equiv 1$, $Y_1(s)\equiv \sgn(s)$ for even and odd parity respectively. Here $\sgn(s)=1$ for $s>0$, $\sgn(s)=0$ for $s=0$, and $\sgn(s)=-1$ for $s<0$. 
$\delta_p$ is the $p$-wave scattering phase shift which obeys the effective range expansion in 1D \cite{hammer2009causality,hammer2010causality}:
\begin{equation}
	k\cot \delta_p=-\frac{1}{a_p}+\frac{1}{2}r_p k^2+\frac{1}{4!}r_p' k^4+O(k^6).
\end{equation}
Here $a_p$ is the $p$-wave scattering length in 1D, $r_p$ is the $p$-wave effective range, and $r_p'$ is the $p$-wave shape parameter.

If the collision energy is small, namely $|k| \ll 1/r_e$, the wave function can be expanded in powers of $k^2$:
\begin{equation}
\Phi^{(k)}(s)=\phi(s)+k^2 f(s)+k^4 g(s)+O(k^6),
\end{equation}
where $\phi,f,g,\dots$ are called the two-body special functions, and they satisfy the equations \cite{wang2021threebody,wang2021fermion3D}:
\begin{equation}
	\widetilde{H}\phi=0,~~\widetilde{H}f=\phi,~~\widetilde{H} g=f,~\dots,
\end{equation}
where $\widetilde{H}$ is defined as 
\begin{equation}\label{Htilde}
    \widetilde{H} \equiv -\frac{d^2}{ds^2}+ \frac{m}{\hbar^2} V(s).
\end{equation}
%We call $l=0$ the $s$-wave channel and call $l=1$ the $p$-wave channel.
The two-body special functions at $|s|>r_e$ can be extracted from \Eq{two-body-Phi}. By choosing the coefficient $A=-a_p/\sin\delta_p$, we get
\begin{subequations}\label{phi-f-g}
\begin{align}
	&\phi(s)=\left(|s|-a_p\right)\sgn(s),\\
	&f(s)=\left(-\frac{|s|^3}{6}+\frac{a_p}{2}|s|^2-\frac{1}{2} a_p r_p |s|\right)\sgn(s),\\
	&g(s)=\left(\frac{|s|^5}{120}-\frac{a_p}{24}|s|^4+\frac{a_p r_p}{12} |s|^3-\frac{a_p r_p'}{24}|s|\right)\sgn(s)
\end{align}
\end{subequations}
for $|s|>r_e$.

\section{ASYMPTOTICS OF THE THREE-BODY WAVE FUNCTION}\label{sec:asymp}

We consider the collision of three fermions with finite range interactions at zero energy in the center-of-mass frame.
The three-body wave function $\Psi(x_1,x_2,x_3)$ satisfies the following Schr\"{o}dinger equation:
\begin{equation}\label{3body_equ}
-\sum_{i=1}^{3}\frac{\hbar^2}{2m}\frac{\partial^2\Psi}{\partial x_i^2}+ \sum_{i=1}^{3}V(s_i)\Psi+U(s_1,s_2,s_3)\Psi=0,
\end{equation}
where $x_i$ is the coordinate of the $i$th fermion, and $s_i\equiv x_j-x_k$. The indices $(i,j,k)=(1,2,3)$, $(2,3,1)$, or $(3,1,2)$.
$U$ is the three-body potential.
We assume that the interactions among these fermions depend only on the interparticle distances.
The total momentum of the three fermions is zero such that the wave function is translationally invariant.
We assume that $V(s_i)=0$ if $|s_i|>r_e$, and that $U(s_1,s_2,s_3)=0$ if $|s_1|$, $|s_2|$, or $|s_3|$ is greater than $r_e$.

To uniquely determine the wave function for the zero energy collision, we need to also specify the asymptotic behavior of $\Psi$ when the three particles are far apart. 
Suppose that the leading-order term $\Psi_0$ in the wave function scales as $B^p$ at large $B$, where $B=\sqrt{(s_1^2+s_2^2+s_3^2)/2}$ is the hyperradius. $\Psi_0$ should also satisfy the free Schr\"{o}dinger equation $(\partial_1^2+\partial_2^2+\partial_3^2)\Psi_0=0$.
The most important channel for zero-energy collisions, for purposes of understanding ultracold collisions, should be the one with the minimum value of $p$ \cite{wang2021fermion3D}.
%The larger the value of $p$, the less likely for the three particles to come to the range of interaction within which they can interact. 
We find that the minimum value of $p$ for three identical fermions in 1D is $p_{\textrm{min}}=3$, and the leading order term $\Psi_0$ is
\begin{equation}\label{leading-order}
	\Psi_0=s_1 s_2s_3=(x_2-x_3)(x_3-x_1)(x_1-x_2).
\end{equation}
One can check that $\Psi_0$ in \Eq{leading-order} is translationally invariant and it obeys the Fermi statistics.

Like what we did in previous works \cite{tan2008three,wang2021threebody,wang2021fermion3D,wang2022fermion2D}, we derive the corresponding 111 expansion and 21 expansion for the three-body wave function $\Psi$.
When the three particles are all far apart from each other, such that the pairwise distances $|s_1|$, $|s_2|$, $|s_3|$ go to infinity simultaneously for any fixed ratio $s_1:s_2:s_3$, we expand $\Psi$ in powers of $1/B$ and this expansion is called the 111 expansion.
When one fermion is far away from the other two, but the two fermions are held at a fixed distance $s_i$, we expand $\Psi$ in powers of $1/R_i$, where $R_i=x_i-(x_j+x_k)/2$ is a Jacobi coordinate, and this is called the 21 expansion.
These expansions can be written as
\begin{subequations}
\begin{align}
		&\Psi=\sum_{p=-3}^{\infty} \T^{(-p)}(x_1,x_2,x_3),\label{111-form}\\
		&\Psi=\sum_{q=-2}^{\infty}\S^{(-q)}(R,s),\label{21-form}
\end{align}
\end{subequations}
where $\T^{(-p)}$ scales like $B^{-p}$, $\mathcal{S}^{(-q)}$ scales like $R^{-q}$,
and $R\equiv R_i$ and $s\equiv s_i$ for any $i$.
$\T^{(-p)}$ satisfies the free Schr\"odinger equation outside of the interaction range:
\begin{equation}\label{T-p}
	-\left( \frac{\partial^2}{\partial s^2}+\frac{3}{4}\frac{\partial^2}{\partial R^2}\right) \T^{(-p)}=0.
\end{equation}
If one fermion is far away from the other two, \Eq{3body_equ} becomes
\begin{equation}
\left(\widetilde{H}-\frac{3}{4}\frac{\partial^2}{\partial R^2}\right)\Psi =0.
\end{equation}
%where $\widetilde{H}\equiv-\frac{\partial^2}{\partial s^2}+\frac{m}{\hbar^2}V(s)$.
where $\widetilde{H}$ is defined in \Eq{Htilde}, but the $\frac{d}{ds}$ should be replaced by $\frac{\partial}{\partial s}$ here.
Therefore, $\mathcal{S}^{(-q)}$ satisfies the following equations,
\begin{align}
    &\widetilde{H} \mathcal{S}^{(2)}=0,\quad\widetilde{H} \mathcal{S}^{(1)}=0,\nonumber\\
    &\widetilde{H} \mathcal{S}^{(-q)}=\frac{3}{4}\frac{\partial^2}{\partial R^2} \mathcal{S} ^{(-q+2)}\quad (q\geqslant 0).
\end{align}
%Because the three-body wave function $\Psi$ may be expanded as $\sum_p \mathcal{T}^{(-p)}$ at $B\to\infty$, and may also be expanded as $\sum_q \mathcal{S}^{(-q)}$ at $R\to\infty$, the wave function has a double expansion in the region $r_e\ll s\ll R$.
%\begin{equation}
    %\Psi=\sum_{i,j}t^{(i,j)}
%\end{equation}
%where $t^{(i,j)}$ scales like $R^i s^j$.
\begin{figure}[htb]
\includegraphics[width=0.5\textwidth]{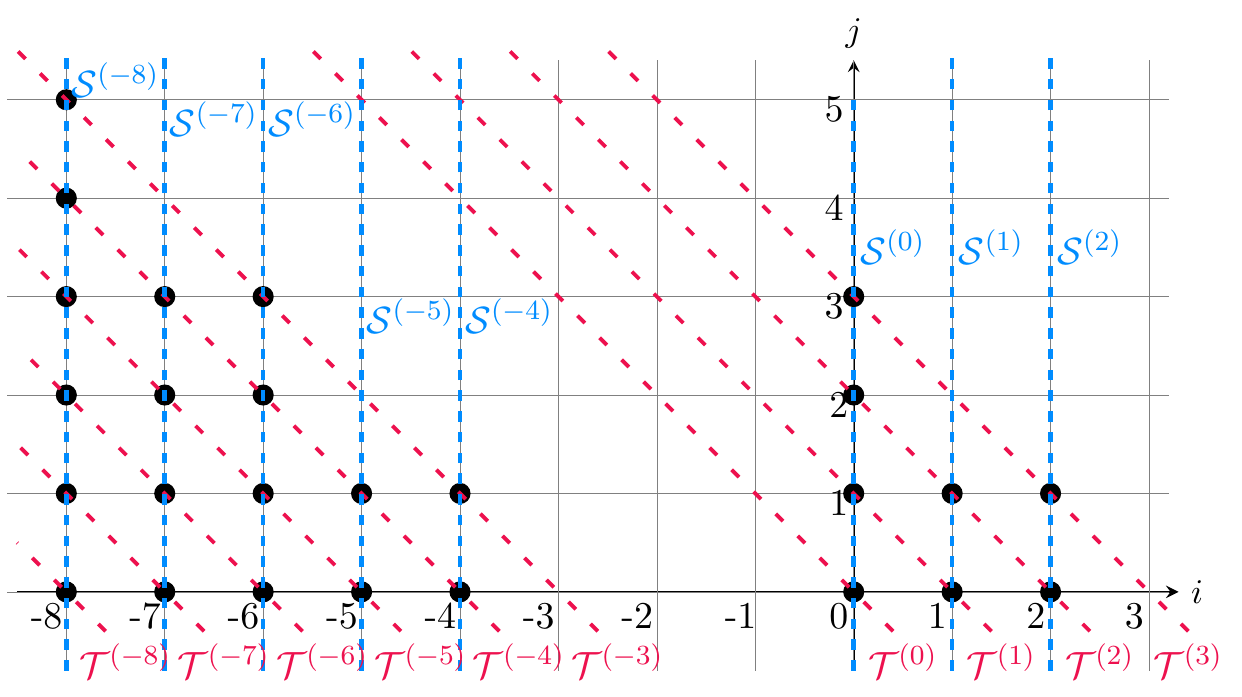}
% Here is how to import EPS art
\caption{\label{fig:expansion} Diagram of the points representing $t^{(i,j)}$ on the $(i,j)$ plane. Each point with coordinates $(i,j)$ represents $t^{(i,j)}$ which scales like $R^i s^j$. Thick dots represent those points at which $t^{(i,j)}\ne0$.
		The term $\mathcal{T}^{(-p)}$ in the 111 expansion is represented by a red dashed line satisfying the equation $i+j=-p$. The term $\mathcal{S}^{(-q)}$ in the 21 expansion is represented by a blue dashed line satisfying the equation $i=-q$.}
\end{figure}

To derive the two expansions, we start from the leading-order term in the 111 expansion (which fixes the overall amplitude of $\Psi$):
\begin{equation}
	\mathcal{T}^{(3)}=\Psi_0=\frac{1}{4}s^3-sR^2.
\end{equation}
We then first derive $\mathcal{S}^{(2)}$, and then derive $\mathcal{T}^{(2)}$, and then derive $\mathcal{S}^{(1)}$, and then derive $\mathcal{T}^{(1)}$, and so on, all the way until $\mathcal{S}^{(-8)}$. At every step, we require the 111 expansion and the 21 expansion to be consistent in the region $r_e\ll |s|\ll |R|$, in which the wave function has a double expansion:
\begin{equation}
\Psi=\sum_{i,j}t^{(i,j)},    
\end{equation}
where $t^{(i,j)}$ scales as $R^i s^j$,
and
\begin{align}
\mathcal{T}^{(-p)}&=\sum_{i}t^{(i,-p-i)},\\
\mathcal{S}^{(-q)}&=\sum_j t^{(-q,j)}.
\end{align}
In Fig.~\ref{fig:expansion} we show the points on the $(i,j)$ plane for which $t^{(i,j)}$
is nonzero.
Our resultant  111 expansion is
\begin{widetext}
\begin{align}
\Psi&=s_1 s_2 s_3\left(1-\frac{3\sqrt{3}D_F}{2\pi B^6}\right)+\sum_{i=1}^{3}\bigg[ -a_p B^2 \cos( 2\Theta_i )\sgn(s_i)-\frac{6}{\pi}a_p^2 B \theta_i\sin\theta_i\sgn(s_i)+\frac{3}{4}\left(2a_p^3+a_p^2 r_p\right) \sgn(s_i)\nonumber\\
&-\frac{3\sqrt{3}\,a_pD_F}{2\pi B^4} \cos (4\Theta_i)\sgn(s_i)-\frac{18\sqrt{3}\,a_p^2D_F}{\pi^2 B^5} \theta_i \sin (5\theta_i) \sgn(s_i)+\frac{45\sqrt{3}D_F}{4\pi B^6}\left(2a_p^3+a_p^2r_p\right)  \cos (6\Theta_i) \sgn(s_i)\nonumber\\
&+\frac{405\sqrt{3}}{2\pi^2 B^7}a_p^3 r_pD_F \theta_i \sin (7\theta_i) \sgn(s_i)-\frac{945\sqrt{3}D_F}{32\pi B^8}\left(6a_p^3r_p^2+a_p^2r_p'\right)  \cos (8\Theta_i) \sgn(s_i) \bigg]+O(B^{-9}),\label{111}
\end{align}
where $D_F$ is the three-body scattering hypervolume.
% of identical spin-polarized fermions in 1D.
The coefficient in $\T^{(-3)}$ is chosen such that
$
    (\partial_s^2+\frac{3}{4}\partial_R^2)\T^{(-3)}=\frac{3}{4} D_F [\delta'(s)\delta''(R)-\frac{4}{9}\delta'''(s)\delta(R)],
$
and this coefficient will simplify the expression for the shift of the energy of three fermions along a periodic line; see \Eq{energy-3fermion}.
$\Theta_i$ is called the hyperangle and is defined via the following equations:
\begin{equation}
	\frac{\sqrt{3}}{2}s_i=B \cos \Theta_i,~~R_i=B\sin\Theta_i.
\end{equation}
One can verify that the three hyperangles satisfy
$\Theta_1=\Theta_2-\frac{2\pi}{3}+2n\pi$,
$\Theta_3=\Theta_2+\frac{2\pi}{3}+2n'\pi$,
where $n$ and $n'$ are integers. We also define the reduced hyperangle 
$
\theta_i \equiv \arctan\frac{2|R_i|}{\sqrt{3} |s_i|},~~\theta_i\in[0,\frac{\pi}{2}].
$
Three fermions in 1D have 6 different sorting orders. If $x_1<x_2<x_3$, the 111 expansion is simplified as
\begin{align}\label{111-B-th}
    \Psi=&\frac{2}{3\sqrt{3}}B^3 \cos (3\theta_2)-2a_p B^2 \cos (2\theta_2)+2\sqrt{3} a_p^2 B \cos \theta_2-\frac{3}{4}\left(2a_p^3+a_p^2 r_p\right)\nonumber\\
    &-\frac{D_F}{\pi B^3}\cos (3\theta_2)-\frac{3\sqrt{3}D_Fa_p}{\pi B^4}\cos (4\theta_2)-\frac{18D_F a_p^2 }{\pi B^5}\cos (5\theta_2)-\frac{45\sqrt{3}D_F}{4\pi B^6}\left(2a_p^3+a_p^2 r_p\right)\cos (6\theta_2)\nonumber\\
    &-\frac{405D_F a_p^3 r_p}{2\pi B^7}\cos (7\theta_2)-\frac{945\sqrt{3}D_F}{16\pi B^8}\left(6a_p^3r_p^2+a_p^2 r_p'\right)\cos (8\theta_2)+O(B^{-9}).
\end{align}

Our resultant 21 expansion is
\begin{align}\label{21}
\Psi=&\Bigg[-R^2+3a_p|R|-\frac{3}{4}\left(2a_p^2+a_p r_p\right)+\frac{3\sqrt{3}D_F}{2\pi R^4}+\frac{9\sqrt{3}\,a_pD_F}{\pi |R|^5}+\frac{45\sqrt{3}D_F}{4\pi R^6}\left(2a_p^2+a_p r_p\right)\nonumber\\
&\quad+\frac{405\sqrt{3}D_F}{4\pi|R|^7} a_p^2 r_p + \frac{945\sqrt{3}D_F}{32\pi R^8} \left(6a_p^2 r_p^2+a_p r_p'\right)\Bigg]\phi(s)\nonumber\\
&+\left[-\frac{3}{2}+\frac{45\sqrt{3}D_F}{2\pi R^6}+\frac{405\sqrt{3}\,a_pD_F}{2\pi|R|^7} +\frac{2835\sqrt{3}D_F}{8\pi R^8} \left(2a_p^2+a_p r_p\right) \right]f(s)+\frac{2835\sqrt{3}D_F}{4\pi R^8} g(s)+O(R^{-9}).
\end{align}

\end{widetext}
We need to emphasize that \Eq{21} is applicable when the interaction does not support any two-body bound states. 
If the interaction supports $n_b$ two-body bound states, three fermions may form such a two-body bound state and a free fermion, which fly
apart with total kinetic energy equal to the released two-body binding energy. In this case, the 21 expansion is modified as \cite{zhu2017threebody}
\begin{equation}\label{21mod}
    \Psi=\Psi_{21}+\sum_{n=1}^{n_b} c_n \phi_n(s) \exp\left(i\frac{2}{\sqrt{3}}\kappa_n |R|\right),
\end{equation}
where $\Psi_{21}$ is defined as the right-hand side of \Eq{21}.
The second term on the right-hand side of \Eq{21mod}
is the outgoing wave with wave number $2\kappa_n/\sqrt{3}>0$. Here $\phi_n$ is the wave function of the $n$th two-body $p$-wave bound state with energy $E_n=-\hbar^2\kappa_n^2/m$, and satisfies the Schr\"{o}dinger equation and the normalization condition:
\begin{align}
    &\left(-\frac{d^2}{ds^2}+\frac{mV(s)}{\hbar^2}+\kappa_n^2\right)\phi_n(s)=0,\\
    &\int_{-\infty}^{\infty} ds~ |\phi_n(s)|^2=1.
\end{align}
The coefficients $c_n$ are in general non-universal parameters that depend on the details of the interaction potentials.
$c_n$ determines the probability amplitude of producing the $n$th bound state which fly apart from the remaining fermion after the three-body zero-energy collision. But using probability conservation, one can show that these coefficients are related to the imaginary part of the
three-body scattering hypervolume.
As the outgoing wave contributes a positive probability flux towards the outside of a large circle centered at the origin in the plane of coordinates $(\frac{\sqrt3}{2}s,R)$, $D_F$ gains a negative imaginary part to make the total flux through the circle vanish and conserve the probability. From this conservation of probability we derive the relation between the imaginary part of $D_F$ and the norm-squares of the coefficients $c_n$:
\begin{equation}
    \mathrm{Im}D_F=-\frac{3\sqrt{3}}{2}\sum_{n=1}^{n_b} \kappa_n |c_n|^2.
\end{equation}
Even if $n_b=1$, one can \emph{not} determine $c_1$ completely from $\im D_F$,
because the phase of $c_1$ can not be determined from $\im D_F$. 
To determine $c_n$, one need to solve the three-body Schr\"{o}dinger equation using
the actual interaction potentials between the fermions.

In Sec.~\ref{sec:recombination} we will study the relation between $\im D_F$ and the three-body recombination rates of one-dimensional ultracold spin-polarized Fermi gases. 

\section{Evaluation of the scattering hypervolume for several interaction potentials}\label{sec:value-DF}
In this section, we first derive an approximate formula for the hypervolume $D_F$ for weak potentials by using the Born expansion. 
We then numerically compute $D_F$ for the square-barrier and the square-well pairwise potentials and the Gaussian pairwise potentials having various strengths.

\subsection{Weak interaction potentials}
If the potentials $V(s)$ and $U(s_1,s_2,s_3)$ are waek, we can express the wave function as a Born expansion \cite{zhu2017threebody,wang2022fermion2D}:
\begin{equation}
\Psi=\Psi_0+\Psi_1+\Psi_2+\cdots,
\end{equation}
where $\Psi_0=s_1s_2s_3=s^3/4-sR^2$ is the wave function of three free fermions, $\Psi_n=(\widehat{G}\mathcal{V})^n\Psi_0$,
$\widehat{G}=-\widehat{H}_0^{-1}$ is the Green's operator, $\widehat{H}_0$ is the three-body kinetic-energy operator, and $\mathcal{V}=U(s_1,s_2,s_3)+\sum_i V(s_i)$ is the interaction potential. 

We derive the first-order and the second-order corrections at $|s_i|\gg r_e$:
\begin{subequations}\label{Born}
\begin{align}
	&\Psi_1= -\frac{3\sqrt{3}s_1s_2s_3}{4\pi B^6}\Lambda -\sum_{i=1}^{3}\left( \alpha_1 B^2\cos 2\Theta_i+\frac{\alpha_3}{2}\right) \sgn(s_i)\nn\\
	&\quad\quad\,\,+O(UB^{-9}),\label{Born1}\\
	&\Psi_2=\sum_{i=1}^{3}\left[ \beta_1 B^2 \cos 2\Theta_i -\frac{6\alpha_1^2}{\pi}R_i \theta_i +\beta_3\right] \sgn(s_i)\nonumber\\
	&\quad\quad~-\frac{3\sqrt{3}s_1s_2s_3}{20\pi B^6}\left(25\alpha_3^2-7\alpha_1\alpha_5\right)+O(V^2B^{-9})\nn\\
	&\quad\quad\,\,+O(UV)+O(U^2),\label{Born2}
\end{align}
\end{subequations}
where
\begin{subequations}
\begin{align}
	&\alpha_n=\frac{m}{\hbar^2}\int_0^{\infty}\!\!\!ds \:s^{n+1}V(s),\label{alpha}\\
	&\beta_1=\frac{m^2}{\hbar^4}\int_0^{\infty}\!\!\!ds\int_0^{s}\!\!\!ds'\: 2s s'^2 V(s)V(s'),\\
	&\beta_3=\frac{m^2}{\hbar^4}\int_0^{\infty}\!\!\!ds\int_0^{s}\!\!\!ds'\: (ss'^4+2s^3 s'^2 ) V(s)V(s'),\\
	&\Lambda=\frac{m}{\hbar^2}\int_{-\infty}^{\infty}\!\!\!ds'\int_{-\infty}^{\infty}\!\!\!dR'\: \Big(\frac{1}{4}s'^3-s' R'^2\Big)^2 U(s',R').
\end{align}
\end{subequations}
See Appendix \ref{sec:Born} for details of the derivation.

By comparing the results in Eqs.~\eqref{Born} with the 111 expansion in \Eq{111}, we find the expansions of $a_p$ and $D_F$ in powers of the interaction potential:
\begin{align}
    a_p&=\alpha_1-\beta_1+O(V^3),\label{ap-Born}\\
    D_F&=\frac{\Lambda}{2}+\frac{1}{10}(25\alpha_3^2-7\alpha_1\alpha_5)+O(V^3)\nonumber\\
    &~~~+O(UV)+O(U^2).\label{DF-Born}
\end{align}
For any particular two-body potential $V(s)$, e.g., the square-well potential or the Gaussian potential, one can calculate $a_p$ by solving the two-body Schr\"{o}dinger equation and verify that the result is consistent with \Eq{ap-Born} if $V$ is weak.
Equation \eqref{DF-Born} shows that $D_F$ is quadratically dependent on the
two-body potential $V$ if $V$ is weak and the three-body potential $U$ is absent. On the other hand, $D_F$ is linearly dependent on $U$ if $U$ is weak.

If the interactions are not weak, one can solve the three-body Schr\"{o}dinger equation numerically at zero energy and match the resultant wave function with the asymptotic expansions in \Eq{111} and \Eq{21mod} to numerically extract the value of $D_F$.

\subsection{Numerical computations}

%For general interaction strengths, we need to numerically solve the three-body Schr\"{o}dinger equation at zero energy and match the resultant wave function with the asymptotic expansions in \Eq{111} or \Eq{21} to numerically extract the values of $D_F$.

%The three-body wave function $\Psi$ in general is a function of $(s,R,R_c)$, where $R_c=(x_1+x_2+x_3)/3$ is the coordinate of the center of mass. 
The three-body problem in 1D for zero total momentum is equivalent to a one-body problem on a 2D plane. 
The three-body wave function $\Psi$ here depends only on $(s,R)$ or $(B,\Theta)$,
where $s\equiv s_2$, $R\equiv R_2$, and $\Theta\equiv\Theta_2$.
We define the two-dimensional vector $\vect{B}=(\frac{\sqrt{3}}{2}s,R)$, and $\Psi=\Psi(\vect{B})$.
The zero energy Schr\"{o}dinger equation is
\begin{equation}\label{SEin2D}
	-\nabla^2 \Psi+\frac{4m}{3\hbar^2}\mathcal{V}\Psi=0,
\end{equation}
where  $\nabla^2$ is the Laplace operator in 2D:
\begin{equation}
	\nabla^2=\frac{1}{B}\frac{\partial}{\partial B}\left(B\frac{\partial}{\partial B}\right)+\frac{1}{B^2}\frac{\partial^2}{\partial \Theta^2}.
\end{equation}

\begin{figure}[htbp]
	\includegraphics[width=0.45\textwidth,height=0.40\textwidth]{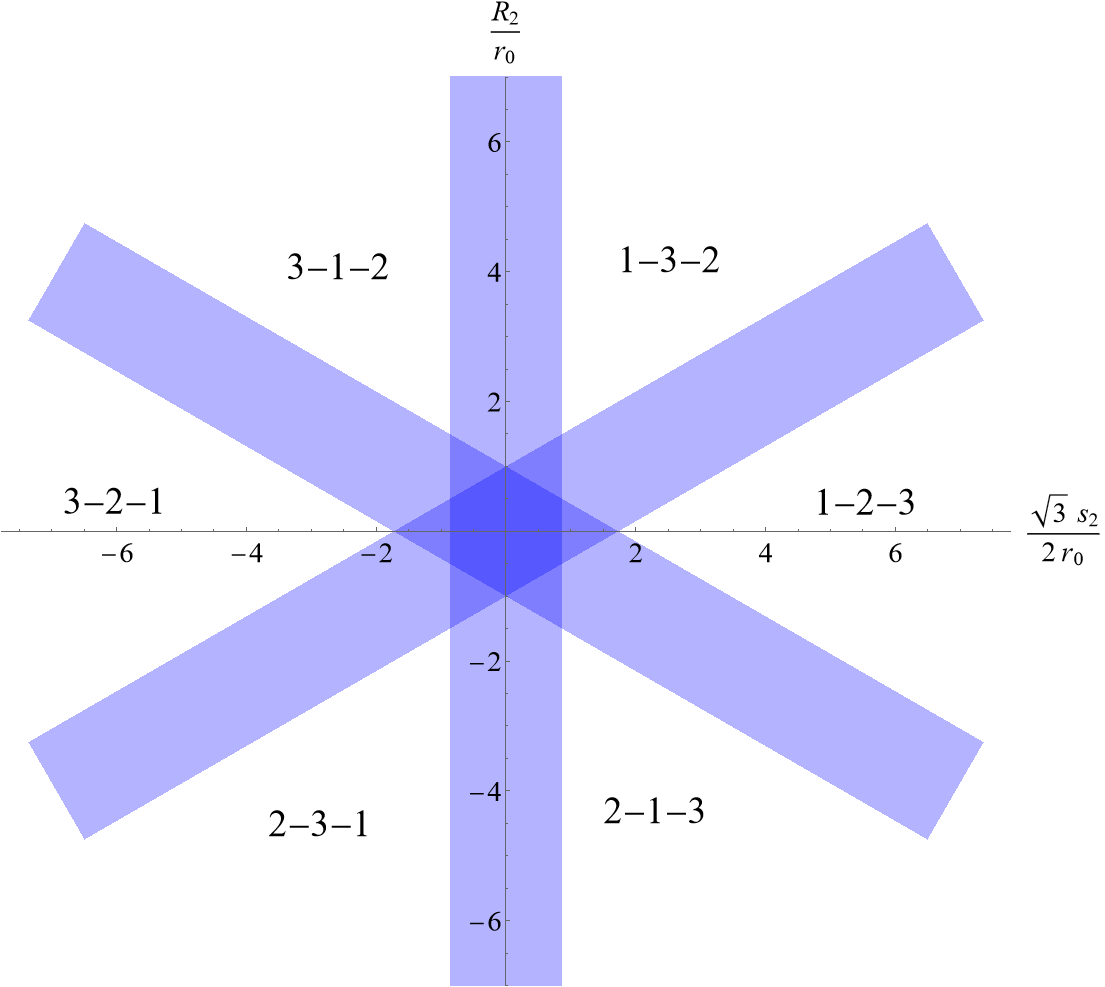}% Here is how to import EPS art
	\caption{\label{fig:2Dplane} The possible configurations of three particles in one dimension. The potential vanishes outside of the colored belts.  The whole plane can be divided into six regions corresponding to six different orders of the coordinates of the three particles. The corresponding order of the three particles is labeled in each region in the figure.}
\end{figure}

Because the interaction potential conserves parity and $\Psi_0$ has odd parity, we can assume that
$\Psi$ has odd parity, namely
\begin{equation}
    \Psi(-x_1,-x_2,-x_3)=-\Psi(x_1,x_2,x_3).
\end{equation}
From the above equation and the Fermi statistics we can show that
\begin{equation}\label{even}
    \Psi(B,-\Theta)=\Psi(B,\Theta)
\end{equation}
and
\begin{equation}\label{periodic}
	\Psi\left( B,\Theta+\frac{\pi}{3}\right)=-\Psi(B,\Theta).
\end{equation}
We can divide the 2D plane into six regions; see Fig.~\ref{fig:2Dplane}. Each region corresponds to a specific order of the coordinates of the three fermions, and we only need to solve \Eq{SEin2D} in one of the six regions.
In the remainder of this section, we always choose to solve the problem in the region $-\pi/6<\Theta<\pi/6$ which corresponds to the order of the coordinates $x_1<x_2<x_3$.

According to \Eq{even} and \Eq{periodic},
the wave function can be expanded as the following Fourier series:
\begin{equation}\label{Psi-fourier}
	\Psi(B,\Theta)=\sum_{i=1}^{\infty} \frac{1}{\sqrt{B}}f_{i}(B)\cos (6i-3) \Theta.
\end{equation}
The potential $\mathcal{V}$ can also be expanded as
\begin{equation}\label{V-fourier}
	\frac{m}{\hbar^2}\mathcal{V}(B,\Theta)=\frac{\nu_0(B)}{2}+\sum_{i=1}^{\infty} \nu_{6i}(B)\cos 6i\Theta.
\end{equation}
The Schr\"odinger equation \eqref{SEin2D} can be written as coupled ordinary differential equations:
\begin{equation}\label{SEvector}
	-f''+\mathcal{U} f=0,
\end{equation}
where $f=(f_1,f_2,f_{3}...)^{\textrm{T}}$ is a column vector, $f''$ means $d^2 f/dB^2$, and $\mathcal{U}=\mathcal{U}(B)$ is a symmetric matrix dependent on $B$.
The matrix elements of $U$ are
\begin{subequations}\label{Umatrix}
	\begin{align}
		&\mathcal{U}_{ii}=\frac{(6i-3)^2-1/4}{B^2}+\frac{2}{3}\left( \nu_0+\nu_{12i-6}\right) ,\\
		&\mathcal{U}_{ij}=\frac{2}{3}\left( \nu_{6|i-j|}+\nu_{6(i+j-1)}\right),~~\textrm{if}~i\neq j 	.
	\end{align}
\end{subequations}
%Note that equations \eqref{Psi-fourier}-\eqref{Umatrix} are valid for any potentials.
Given the wave function on a circle with radius $B$ centered at the origin in the $(\frac{\sqrt3}{2}s_2,R_2)$ plane,
one can use the Schr\"{o}dinger equation to uniquely determine the wave function inside such a circle,
and therefore can determine the partial derivative of the wave function with respect to $B$ on the circle.
Therefore the partial derivative of the wave function with respect to $B$ on such a circle depends linearly
on the wave function on such a circle. So there is a matrix $F$ such that
\begin{equation}\label{def-of-F}
	f'=F f.
\end{equation}
Substituting the above equation into \Eq{SEvector}, and requiring that \Eq{SEvector} be satisfied for all $f$, we find that
 $F$ satisfies a first-order differential equation:
\begin{equation}\label{eq-F}
	F'=\mathcal{U}-F^2.
\end{equation}
At small $B$, we can solve \Eq{SEvector} to find the analytical solution to $f_{i}$ (for square well potentials) or find an expansion of
$f_{i}$ in powers of $B$ (for other potentials); from these we can analytically determine $F$ at infinitesimal $B$, and see that it is diagonal. 
Using the result of $F$ at infinitesimal $B$ as our initial condition, 
we then solve \Eq{eq-F} numerically and determine $F$ at $B=B_0$ for some large $B_0$.
Matching \Eq{def-of-F} at $B=B_0$ with the 111 and the 21 expansions of $\Psi$, we can approximately determine $D_F$.
We then compare the approximate values of $D_F$ determined in this way, using various large values of $B_0$.
We approximately extrapolate to the $B_0\to\infty$ limit
to find the value of $D_F$ with some numerical uncertainty.

\subsubsection{Square-barrier and square-well potentials}
For the square-barrier or square-well potential with strength $V_0$ ($V_0$ can be positive or negative),
\begin{equation}\label{squarewell}
	V(s)=V_0 \frac{\hbar^2}{mr_0^2}\times
	\begin{cases}
		1, & |s|<r_0,\\
		0, & |s|>r_0.
	\end{cases}
\end{equation}
%\textcolor{red}{In the remainder of this subsection, we set $\hbar=m=r_0=1$ for simplicity.}
We can analytically calculate all the Fourier components of $\mathcal{V}$,
\begin{equation}
    \nu_{0}=\frac{V_0}{r_0^2}\times
	\begin{cases}
		6, & 0\leqslant B\leqslant\frac{\sqrt{3}}{2}r_0,\\
		\frac{12}{\pi}\theta_0, & \frac{\sqrt{3}}{2}r_0<B,
	\end{cases}
\end{equation}
\begin{equation}
    \nu_{6i}=\frac{V_0}{{r_0^2}}\times
	\begin{cases}
		0, & 0\leqslant B\leqslant\frac{\sqrt{3}}{2}r_0,\\
		(-1)^i\frac{12}{\pi}\frac{\sin 6i\theta_0}{6i}, & \frac{\sqrt{3}}{2}r_0<B,
	\end{cases}
\end{equation}
for $i\geqslant 1$, where $\theta_0=\arcsin(\sqrt{3}r_0/2B)$.

In the region $B\leqslant\sqrt{3}r_0/2$, the potential $\mathcal{V}=3V_0{\hbar^2/m r_0^2}$ is a constant, and $\nu_0=6V_0/r_0^2$, $\nu_{6i}=0$ for $i\geqslant1$.
So $\mathcal{U}$ is diagonal in this region and $f$ can be analytically determined:
\begin{equation}\label{f-ana}
	f_{i}=\begin{cases}
		c_{i}\sqrt{B}I_{6i-3}(2\sqrt{V_0}B/r_0), & V_0>0,\\
		c_{i}'\sqrt{B}J_{6i-3}(2\sqrt{-V_0}B/{r_0}), & V_0<0,
	\end{cases}
\end{equation}
where $I_j$ is the modified Bessel function of the first kind, and $J_j$ is the Bessel function of the first kind.

At $0<B\leqslant\sqrt{3}{r_0}/2$, $F$ is diagonal and its elements can be easily calculated by using \Eq{f-ana}.
Equation~\eqref{eq-F} is a first-order ordinary differential equation, and the initial value of $F$ at $B=\sqrt{3}{r_0}/2$ is known, so we can compute $F$ numerically at any $B>\sqrt{3}{r_0}/2$.
At large $B$, we use the 111 and the 21 expansions of the wave function in \Eq{111} and \Eq{21mod} to determine $f_1,f_2,f_{3},...$ approximately. By solving \Eq{def-of-F}, we get the numerical value of the scattering hypervolume $D_F$.

\begin{figure}[htbp]
	\includegraphics[width=0.5\textwidth]{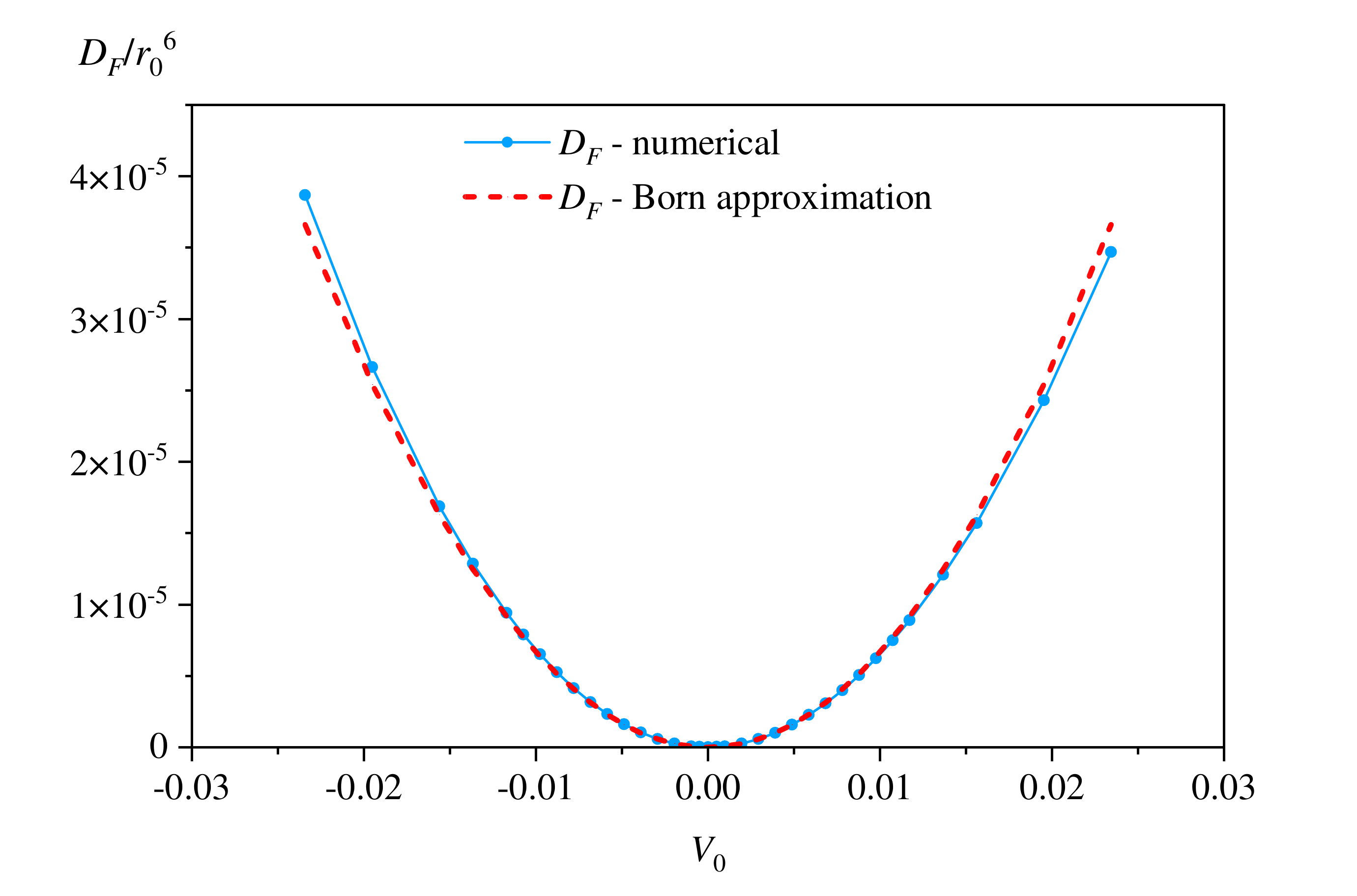}% Here is how to import EPS art
	\caption{\label{fig:square-born} $D_F$ for weak square-barrier or square-well potentials. The blue solid line shows the numerical results and the red dashed line shows the Born approximation.}
\end{figure}
Fig.~\ref{fig:square-born} shows our results of $D_F$ at small $V_0$. 
According to \Eq{alpha} we have
\begin{align}
    		\alpha_n=\frac{V_0}{n+2} r_0^n.
\end{align}
If $V_0$ is small, by using \Eq{DF-Born} we get
\begin{equation}\label{DF-Born-square}
	D_F=\frac{1}{15}V_0^2r_0^6+O(V_0^3).
\end{equation}
The blue solid line in Fig.~\ref{fig:square-born} shows the numerical results and the red dashed line corresponds to the Born approximation $D_F\simeq\frac{1}{15}V_0^2r_0^6$.
The numerical results agree quite well with the Born approximation for small values of $V_0$.

\begin{figure}[htbp]
	\includegraphics[width=0.5\textwidth]{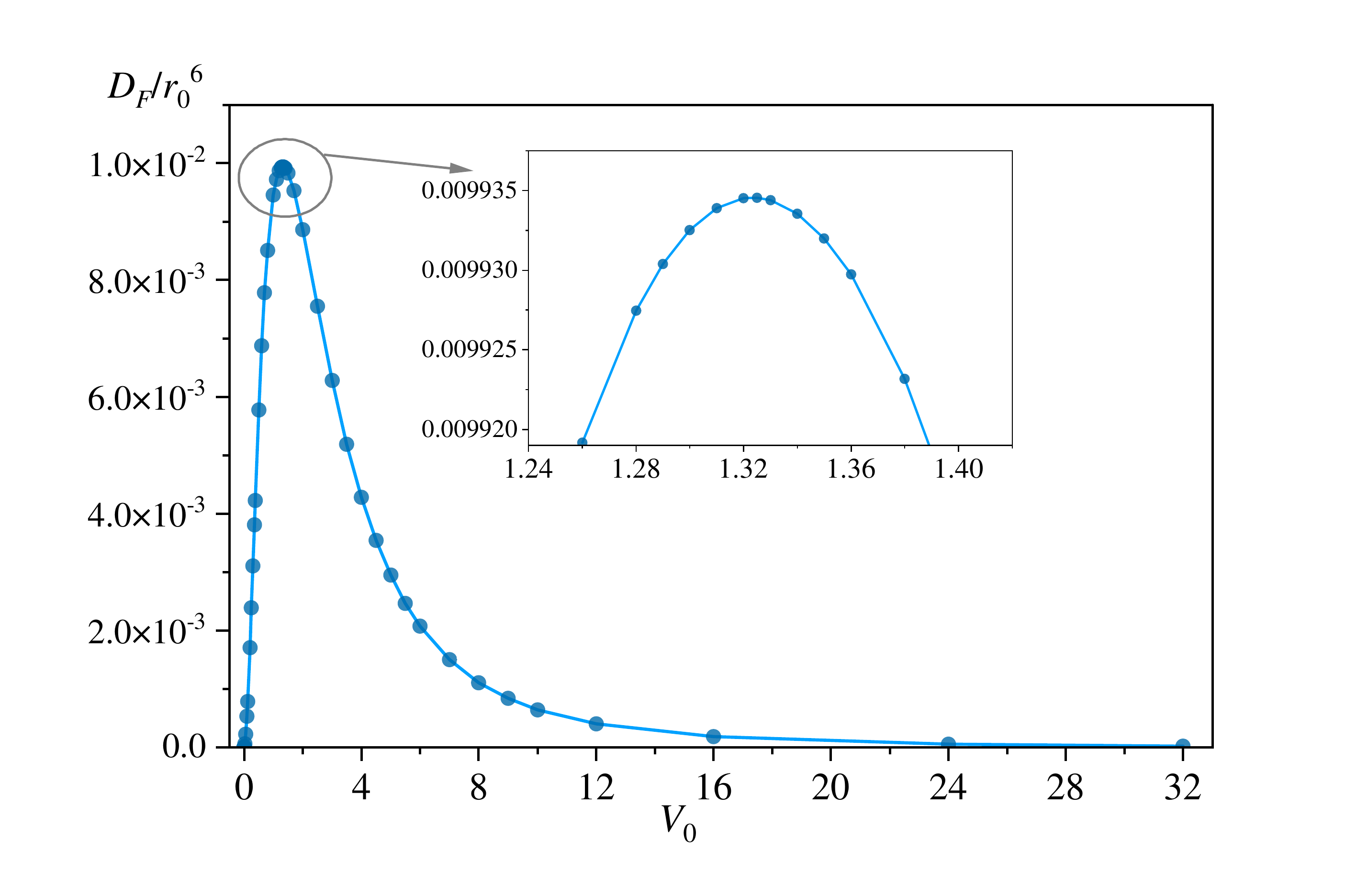}% Here is how to import EPS art
	\caption{\label{fig:square-Df-rep} The value of $D_F$ for the repulsive square-barrier potential defined in \Eq{squarewell}.}
\end{figure}

Fig.~\ref{fig:square-Df-rep} shows the full curve of $D_F$ for repulsive $V_0$.
%%%%%%%%%%%%%%%%%%%%%%%%%%%%%%%%%%%%%%%%%%%%%%%%%%%%%%%%%%%%%%%%%%%%%%%%%%%%%%%%%%%%%
%%%%%%%%%%%%%%%%%%%%%%%%%%%%%%%%%%%%%%%%%%%%%%%%%%%%%%%%%%%%%%%%%%%%%%%%%%%%%%%%%%%%%
$D_F$ increases at $0<V_0<V_c$ where $V_c\simeq1.325$. At $V_0=V_c$, $D_F$ has a maximum of about $0.0099r_0^6$. 
$D_F$ decreases at $V_0>V_c$. 
In the following we will prove that $D_F$ approaches zero as $V_0\rightarrow +\infty$ and scales as $1/V_0^3$ at large $V_0$ for the square-barrier potentials.

If $V_0= +\infty$, the square-barrier potential becomes the hard-core potential. In this case,
the wave function goes to zero in the blue banded region in Fig.~\ref{fig:2Dplane}. We use the new coordinates $\vect{B}'=(\frac{\sqrt{3}}{2} (s-2{r_0}),R)$. 
$\Psi(\vect B)\equiv\widetilde{\Psi}(\vect{B}')$ satisfies the Laplace equation in the sector area, and $\widetilde{\Psi}(\vect{B}')$ satisfies the following boundary conditions:
\begin{equation}
	\widetilde{\Psi}\left(B',\Theta'=-\frac{\pi}{6}\right)=\widetilde{\Psi}\left(B',\Theta'=\frac{\pi}{6}\right)=0.
\end{equation}
where $B',\Theta'$ are defined via
$
	\frac{\sqrt{3}}{2}(s-2{r_0})=B' \cos \Theta',~~R=B'\sin\Theta'.
$
In the domain $-\pi/6<\Theta'<\pi/6$, one can easily find the analytical solution:
\begin{align}\label{HSsolution}
	\widetilde{\Psi}(\vect{B}')=\frac{2}{3\sqrt{3}}B'^3 \cos 3\Theta'.
\end{align}
If we change back to the coordinates $\vect{B}=(\frac{\sqrt{3}}{2}s,R)$, we get
\begin{equation}\label{HSsolution1}
\Psi=\frac{2}{3\sqrt{3}} B^3\cos3\theta_2-2B^2{r_0}\cos 2\theta_2+2\sqrt{3}B{r_0^2}\cos\theta_2-2{r_0^3}.
\end{equation}
Note that at $V_0=+\infty$ \Eq{HSsolution1} is the exact solution and is not just the asymptotic expansion of $\Psi$.
On the other hand, the 111 expansion in this area is simplified as \Eq{111-B-th}.
% \begin{align}\label{111-Bt}
% 	\Psi&=\frac{2}{3\sqrt{3}}B^3 \cos 3\theta_2-2a_p B^2 \cos 2\theta_2+2\sqrt{3} a_p^2 B \cos \theta_2\nonumber\\
% 	&-\frac{3}{2}\left(a_p^3+\frac{1}{2}a_p^2 r_p\right)-\frac{D_F}{\pi B^3}\cos 3\theta_2+\O(B^{-4}).
% \end{align}
%\begin{align}\label{111-Bt}
%	\Psi&=\frac{2}{3\sqrt{3}}B^3 \cos 3\theta_2-2a_p B^2 \cos 2\theta_2+2\sqrt{3} a_p^2 B \cos \theta_2-\frac{3}{2}\left(a_p^3+\frac{1}{2}a_p^2 r_p\right)\nonumber\\
%	&+D_F\Bigg[-\frac{1}{\pi B^3}\cos 3\theta_2-\frac{3\sqrt{3} a_p}{\pi B^4}\cos 4\theta_2-\frac{18 a_p^2}{\pi B^5}\cos 5\theta_2-\frac{45\sqrt{3}}{2\pi B^6}\left(a_p^3+\frac{1}{2}a_p^2 r_p\right)\cos 6\theta_2\nonumber\\
%	&\quad\quad-\frac{405}{2\pi B^7}a_p^3 r_p\cos 7\theta_2-\frac{2835\sqrt{3}}{8\pi B^8}\left(a_p^3r_p^2+\frac{1}{6}a_p^2 r_p'\right)\cos 8\theta_2\Bigg]
%\end{align}
For the hard-core potential with $r_0=1$, we have $a_p={r_0}$, $r_p=2{r_0}/3$. One can check that \Eq{HSsolution1} agrees with \Eq{111-B-th} if $D_F=0$. 
So $D_F=0$ for the hard-core potential, and this is consistent with our numerical results
for the values of $D_F$ for the square-barrier potential at $V_0\to\infty$.

If $V_0$ is large but finite, we also get an expansion in powers of $1/V_0$:
\begin{widetext}
\begin{align}
	&\widetilde{\Psi}(\vect{B}')=\frac{2}{3\sqrt{3}}B'^3 \cos 3\theta'+\frac{2}{\sqrt{V_0}}B'^2\cos 2\theta'+\frac{2\sqrt{3}}{V_0}B'\cos\theta'+\frac{9}{4V_0^{3/2}}+\O(B'^{-3})\nonumber\\
	&=\frac{2}{3\sqrt{3}} B^3\cos3\theta_2-2B^2\cos 2\theta_2\left( 1-\frac{1}{\sqrt{V_0}}\right) +2\sqrt{3}B\cos\theta_2\left( 1-\frac{1}{\sqrt{V_0}}\right)^2-\left( 2-\frac{6}{\sqrt{V_0}}+\frac{6}{V_0}-\frac{9}{4V_0^{3/2}}\right) +\O(B^{-3}).\label{largeV0}
\end{align}
\end{widetext}

\begin{figure}[htb]
\subfloat[]
{	
	\includegraphics[width=0.25\textwidth]{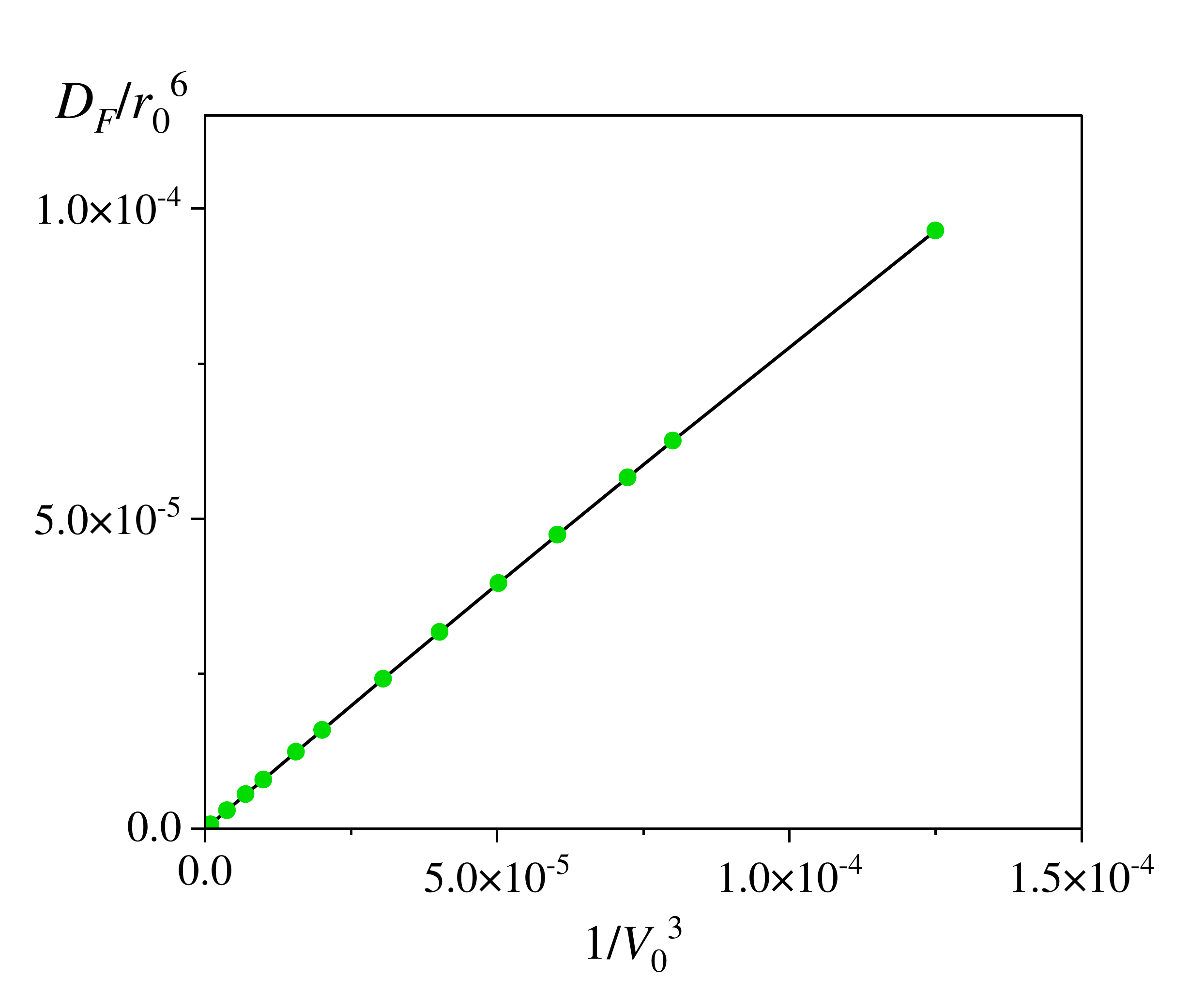}   
}
\subfloat[]
{	
	\includegraphics[width=0.25\textwidth]{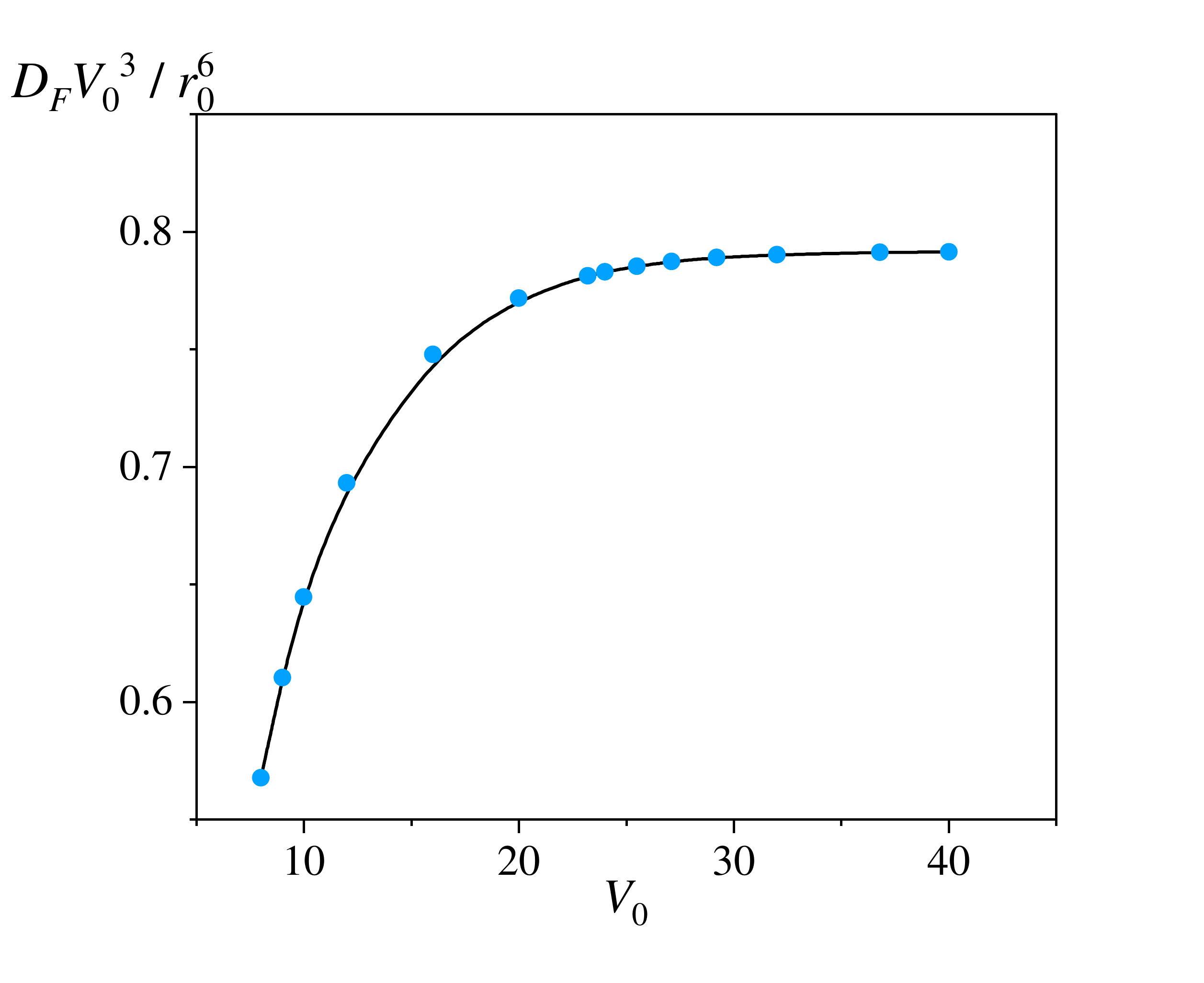} 
}	
\caption{(a) $D_F$ vs $1/V_0^3$ for the repulsive square-barrier potentials. (b) $D_F V_0^3$ vs $V_0$ for these potentials.
The subfigures (a) and (b) both show that $D_F$ is proportional to $1/V_0^3$ if $V_0$ is large.}
\label{fig:DF-V03}
\end{figure}

If $1/\sqrt{V_0}\ll B'\ll r_0$, the wave function $\widetilde{\Psi}(\vect{B}')$ satisfies a scaling law: if $\widetilde{\Psi}(\vect{B}')$ is the solution at interaction strength $V_0$, then $\widetilde{\Psi}(\sqrt{\lambda}\vect{B}')$ is the solution at interaction strength $\lambda V_0$.
According to this we know the next term in the first line of \Eq{largeV0} should take the form $1/V_0^3 B'^3$, which implies that $D_F$ scales as $V_0^{-3}$ at large $V_0$:
\begin{equation}
	D_F=\frac{\mathcal{C}}{V_0^3}+o(V_0^{-3}).
\end{equation} 
Fig.~\ref{fig:DF-V03} shows that our numerical results agree with this.
From the numerical results we get $\mathcal{C}\simeq 0.79$.

\subsubsection{Gaussian potential}
In this subsection we consider the Gaussian potential
\begin{equation}
	V(s)=V_0 \frac{\hbar^2}{m r_0^2} e^{-s^2/r_0^2},
\end{equation}
where the strength $V_0$ can be positive or negative.
According to \Eq{alpha} we get
\begin{align}
		\alpha_n=\frac{1}{2}\Gamma\left(1+\frac{n}{2}\right)V_0 r_0^n.
\end{align}
If $V_0$ is small, by using \Eq{DF-Born} we get the Born approximation of $D_F$:
\begin{equation}\label{DF-Born-gaussian}
	D_F=\frac{3\pi}{16}V_0^2 r_0^6+O(V_0^3).
\end{equation}

To numerically compute the value of $D_F$, we also Fourier-expand the wave function and the potential function.
The Fourier components of $\mathcal{V}$ for the Gaussian potential can be calculated analytically:
\begin{equation}
	\nu_{6i}(B)= (-1)^i \frac{6V_0}{{r_0^2}}  I_{3i}\Big(\frac{2B^2}{3r_0^2}\Big)e^{-2B^2/3r_0^2},
\end{equation}
where $i=0,1,2,\dots$.
At small $B$, unlike the case of square-well potential, we can not get an analytical expression for the matrix $F$ for the Gaussian potential. However we can solve \Eq{SEvector} to find an expansion of $f_{i}$ in powers of $B$, and get an approximate expression for the matrix $F$ at small $B$. The  remaining algorithm is similar to the case of square-well potential, and we get the numerical values of $D_F$ for the repulsive and the attractive Gaussian potentials. 

\begin{figure}[htbp]
	\includegraphics[width=0.5\textwidth]{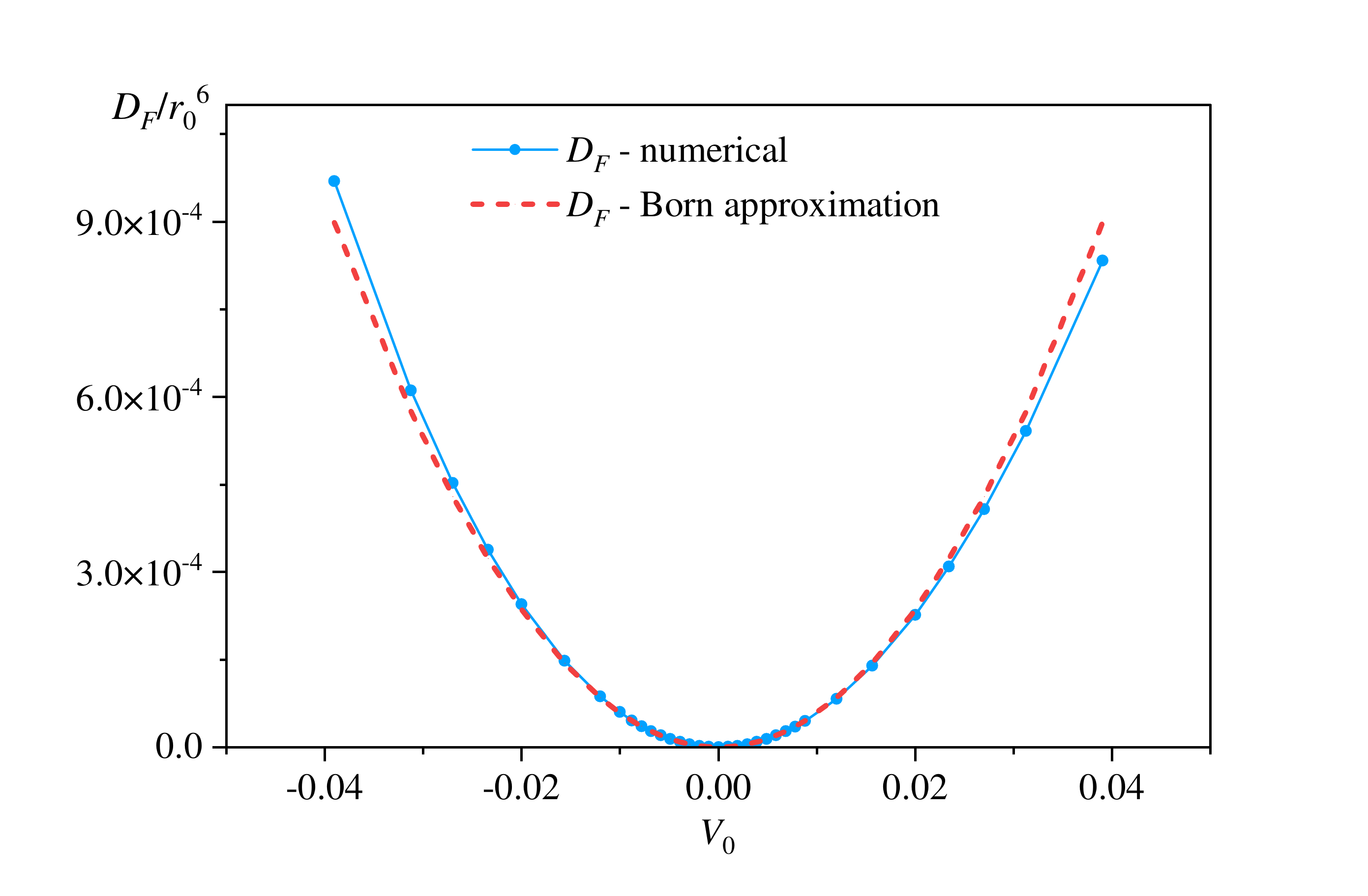}% Here is how to import EPS art
	\caption{\label{fig:gaussian-born} The values of $D_F$ for weak Gaussian potentials. The blue solid line shows the numerical results and the red dashed line shows the Born approximation.}
\end{figure}

Fig.~\ref{fig:gaussian-born} shows our numerical results of $D_F$ at small $V_0$. We see that the results are consistent with the Born approximation.

\begin{figure}[htbp]
	\includegraphics[width=0.5\textwidth]{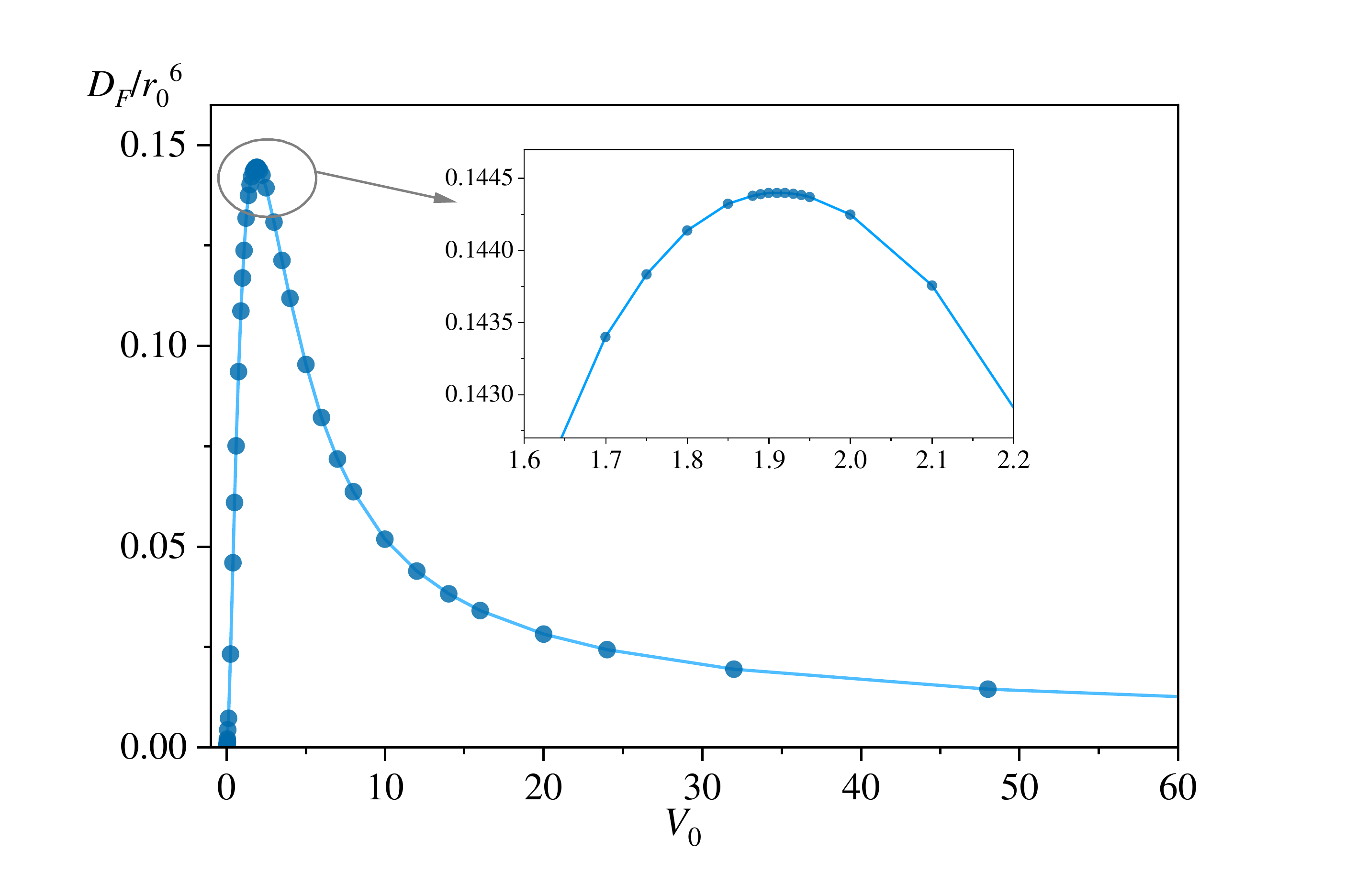}
	\caption{\label{fig:gaussian-rep} The values of $D_F$ for repulsive Gaussian potentials.}
\end{figure}

Fig.~\ref{fig:gaussian-rep} shows the values of $D_F$ for repulsive Gaussian potentials. $D_F/r_0^6$ has a maximum of about $0.144 $ at $V_0\simeq1.91$. $D_F/r_0^6$ decreases at $V_0>1.91$. The rate of the decrease is slower than in the case of square-well  potentials.

\begin{figure*}[htbp]
\centering
	\includegraphics[width=0.95\textwidth]{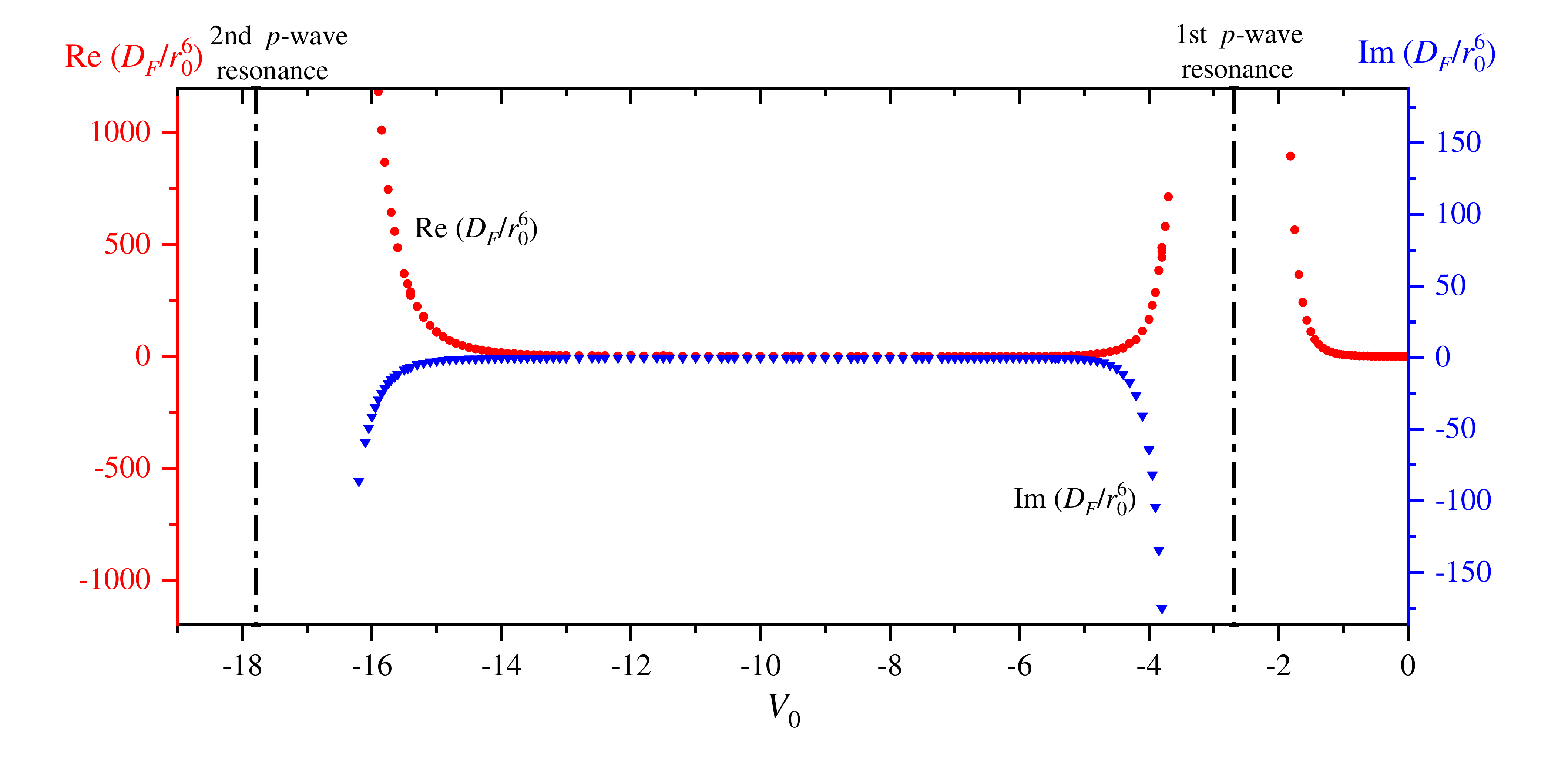}% Here is how to import EPS art
	\caption{\label{fig:gaussian-attr} The values of $D_F$ for attractive Gaussian potentials. The red dots represent the real part of $D_F/r_0^6$ and the blue triangles represent the imaginary part of $D_F/r_0^6$. The vertical dashed lines show the critical strengths of the Gaussian potential at which the $p$-wave resonances occur. }
\end{figure*}

\begin{figure*}[htb]
\subfloat[]
{	
	\includegraphics[width=0.5\textwidth]{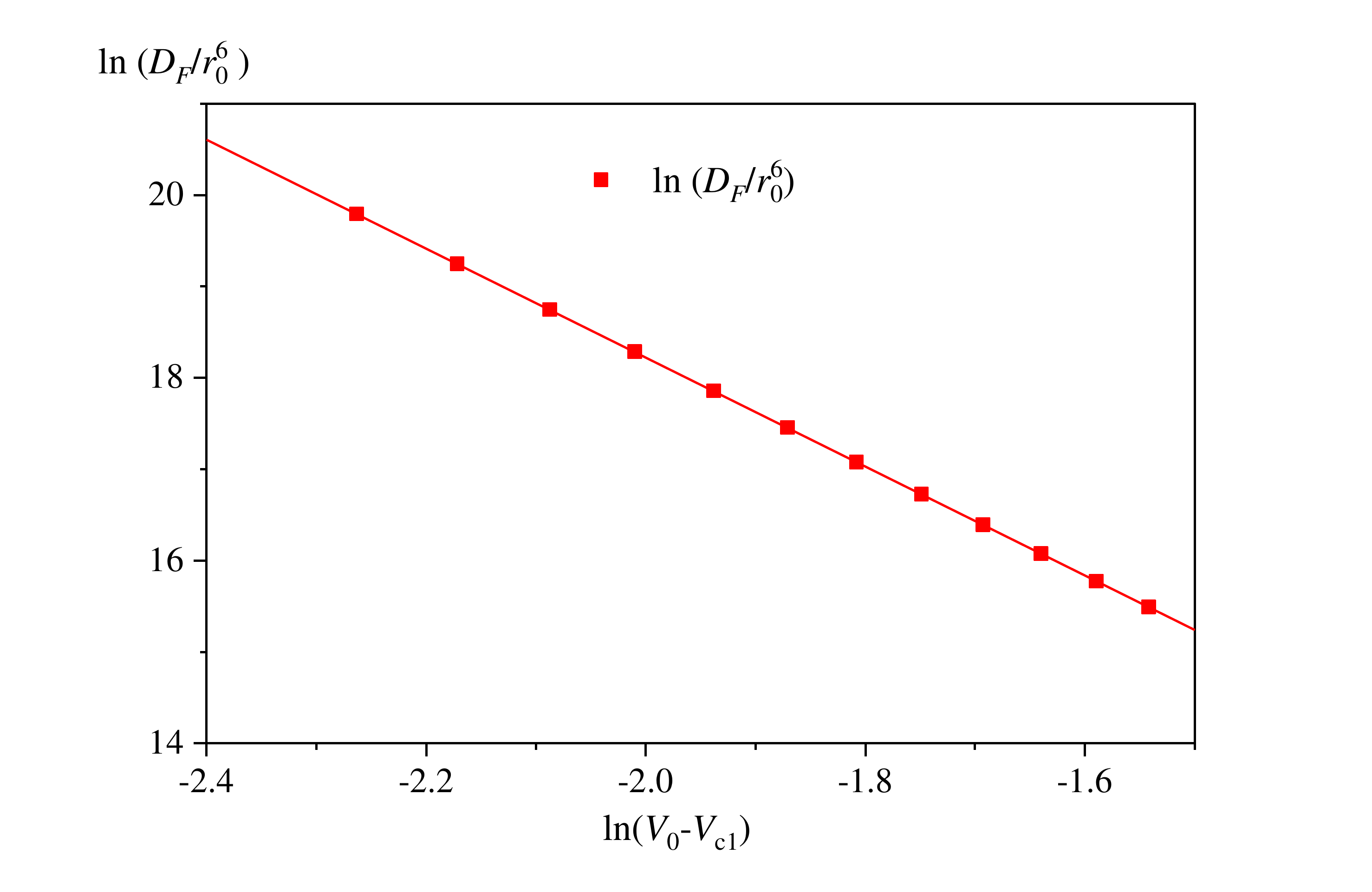}   
	\label{fig:fitB}
}
\subfloat[]
{	
	\includegraphics[width=0.5\textwidth]{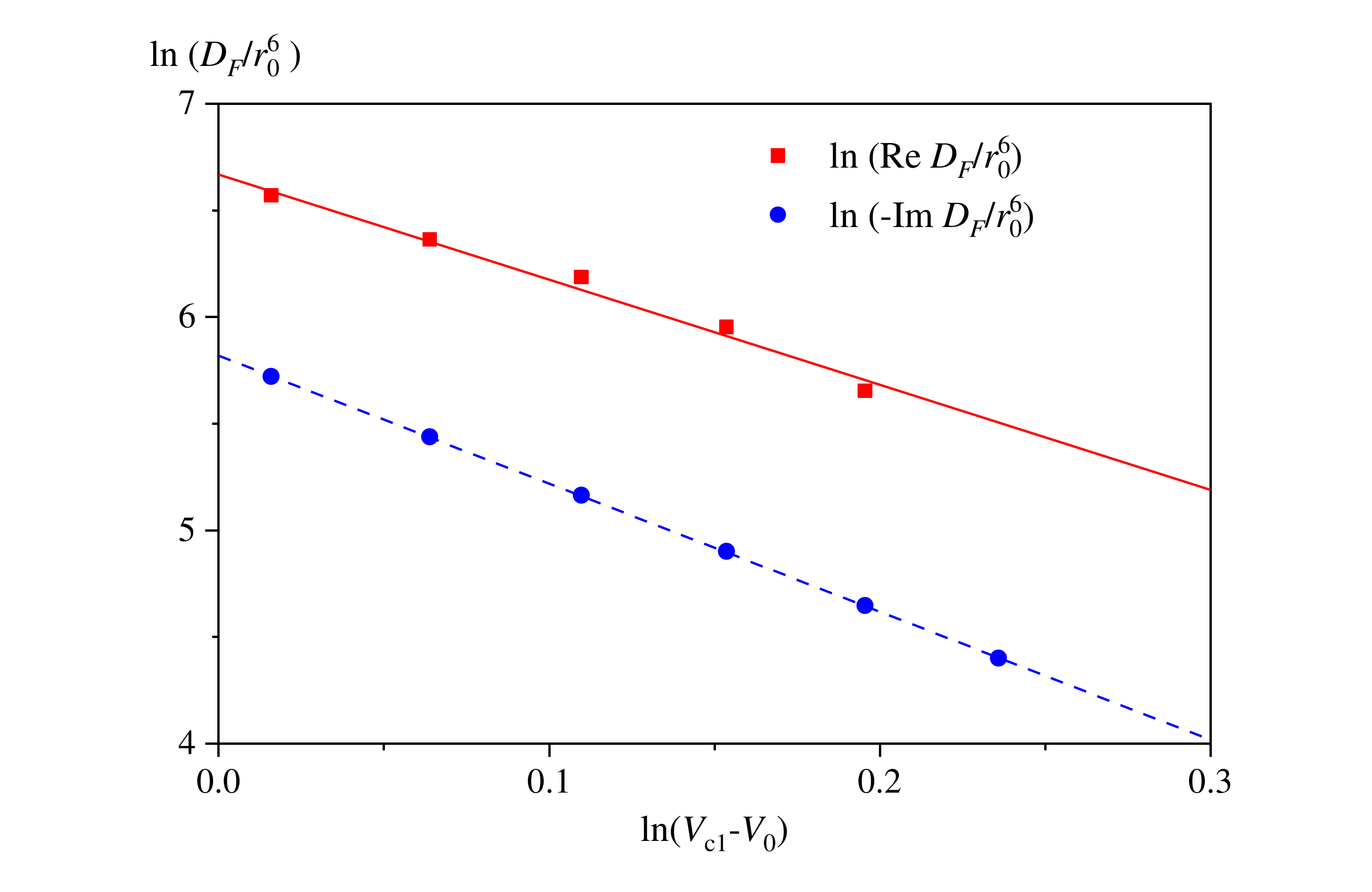} 
	\label{fig:fitA}
}	
\caption{
    (a) $\ln(D_F/r_0^6)$ vs. $\ln(V_0-V_{c1})$ when $V_0$ is slightly greater than $V_{c1}$.
	Doing a linear fit in this double-log plot, we find that
	$D_F\simeq\frac{542.9r_0^6}{(V_0-V_{c1})^{5.963\pm0.002}}\simeq0.74 a_p^6$.
	(b)
	$\ln[\re(D_F/r_0^6)]$ (red squares) and $\ln[-\im(D_F/r_0^6)]$ (blue dots)
	ploted against $\ln(V_{c1}-V_0)$ when $V_0$ is slightly less than $V_{c1}$. Doing linear fits in these double-log plots, we find that
	$ \textrm{Re} D_F\simeq\frac{796{r_0^6}}{(V_{c1}-V_0)^{4.98\pm0.36}}$ and
	$\textrm{Im} D_F\simeq-\frac{337r_0^6}{(V_{c1}-V_0)^{6.00\pm0.02}}\simeq-0.46 a_p^6$
	, where we have used the approximate formula \Eq{ap-resonance}.
	}
\end{figure*}

Fig.~\ref{fig:gaussian-attr} shows our results of $D_F$ for attractive Gaussian potentials. 
If the potential strength is weak, there is no two-body bound state. As the depth of the potential increases, two-body bound states appear one by one. At $V_0=V_{c1}\simeq-2.684$ the first $p$-wave resonance occurs, and the first two-body bound state appears. When $V_0$ is close to $V_{c1}$ we find an approximate formula for $a_p/r_0$:
\begin{equation}\label{ap-resonance}
    a_p/r_0\simeq-3.007/(V_0-V_{c1})+1.041.
\end{equation}
At $V_0=V_{c2}\simeq-17.796$, the second $p$-wave resonance occurs, and the second two-body bound state appears.
These resonances are indicated by the vertical black dot-dashed lines in Fig.~\ref{fig:gaussian-attr}.
At $V_{c1}<V_0<0$ there is no two-body bound state and $D_F$ is real.  When $V_0$ approaches $V_{c1}$ from above, $D_F$ diverges. 
To understand the behavior of $D_F$ when $V_0$ is close to $V_{c1}$, we plot $\ln(D_F/r_0^6)$ vs. $\ln(V_0-V_{c1})$ when $V_0$ is slightly greater than $V_{c1}$, in Fig.~\ref{fig:fitB}. It seems that there is a linear relationship. Doing a linear fit, we find that $D_F$ is proportional to $(V_0-V_{c1})^{-6}$,
and we derive an approximate formula: $D_F\simeq 0.74 a_p^6$ when $V_0$ is slightly greater than $V_{c1}$.

At $V_{c2}<V_0<V_{c1}$ there is one two-body $p$-wave bound state, and in this case $D_F$ gains a negative imaginary part, $D_F=\textrm{Re} D_F+i \textrm{Im} D_F$. 
The absolute value of $\textrm{Im} D_F$ is smaller than the absolute value of $\re D_F$ for most values of $V_0$ in this range. 
When $V_0$ approaches $V_{c1}$ from below, $\re D_F$ and $\im D_F$ both diverge. 
We plot $\ln[\re(D_F/r_0^6)]$ and $\ln[-\im(D_F/r_0^6)]$ vs. $\ln(V_{c1}-V_0)$ when $V_0$ is slightly less than $V_{c1}$, in Fig.~\ref{fig:fitA}. We again see approximately linear relationships. Doing linear fits, we find
that $\re D_F$ seems to be proportional to $(V_{c1}-V_0)^{-5}$ but $\im D_F$ is perhaps proportional to $(V_{c1}-V_0)^{-6}$, and we get an approximate formula: $\im D_F\simeq-0.46 a_p^6$ when $V_0$ is slightly less than $V_{c1}$.
According to the results in Sec.~\ref{sec:recombination}, the divergence of $\im D_F$ indicates that a one-dimensional spin-polarized Fermi gas will suffer strong three-body recombination losses near such resonances.

\begin{figure}[htbp]
\centering
	\includegraphics[width=0.5\textwidth]{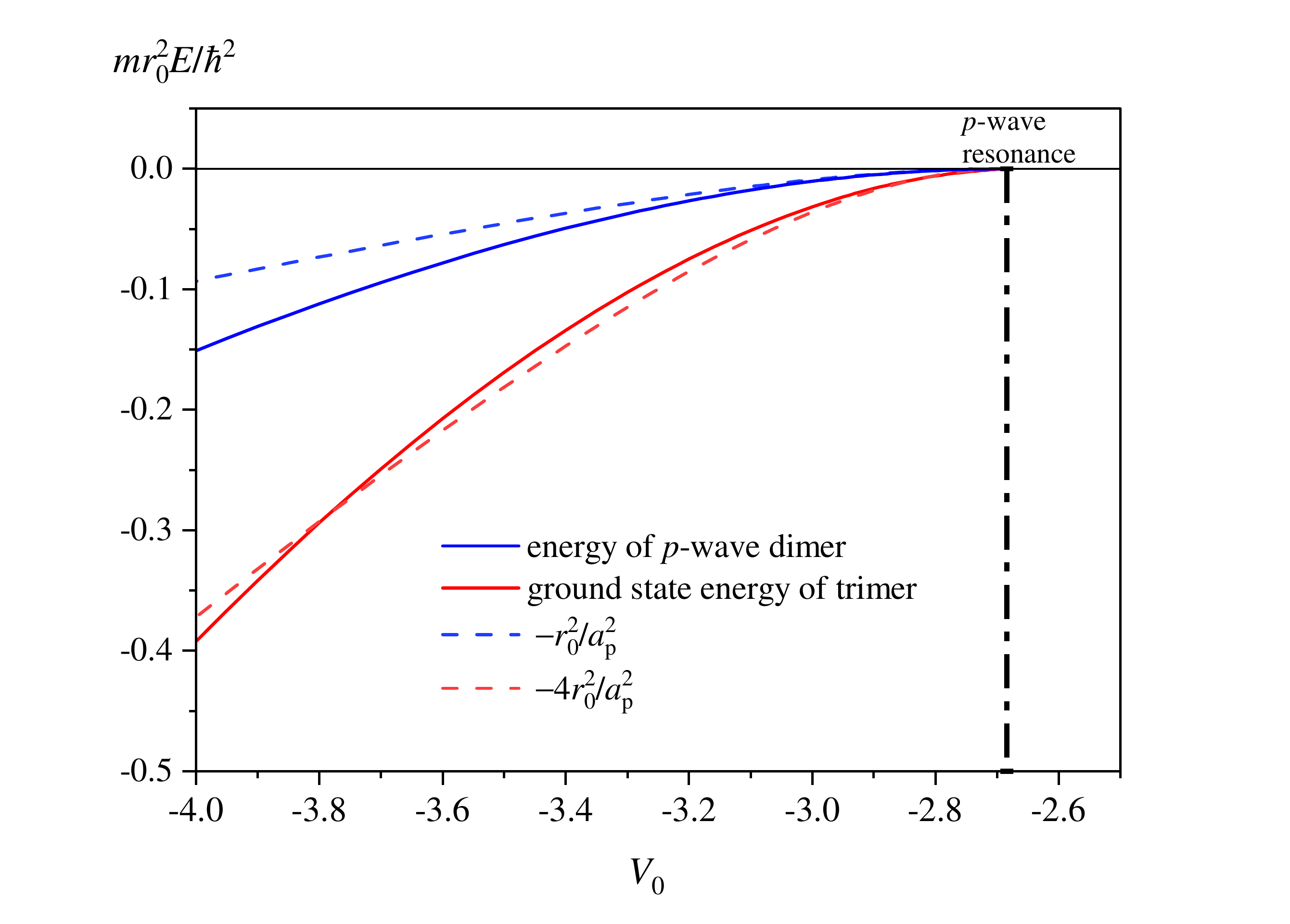}
	\caption{\label{fig:3bodybound}The energies of bound states with Gaussian pairwise interactions vs. the interaction strength $V_0$. The blue solid line shows the two-body bound state energy, and the red solid line shows the three-body bound state energy. 
	The dashed lines correspond to the universal formulas in Eqs.~\eqref{E2universal} and \eqref{E3universal}.}
\end{figure}

%\textcolor{red}{
If $V_0$ is slightly less than $V_{c1}$, $a_p$ is positive and very large,  and the two-body bound state is very shallow. The energy of the shallow bound state satisfies the universal formula:
\begin{equation}\label{E2universal}
    E_2\simeq-\hbar^2/ma_p^2.
\end{equation}
According to the Bose-Fermi duality \cite{Girardeau1960,Fermion-Boson-duality}, the properties of the one-dimensional Fermi system with large and positive scattering length are similar to those of a weakly attractive bosonic system, which can be described by using the Lieb-Liniger model \cite{LiebLiniger1963} with the repulsive contact interaction replaced by attractive contact interaction, and this model can be exactly solved by using the Bethe ansatz \cite{Bethe1931}. Ref~\cite{McGuire1964} shows that such a bosonic system has a three-body bound state with energy $E_3=4E_2$. Mapping this bosonic system to the fermionic system with two-body $p$-wave scattering length $a_p\gg r_0$,
we infer a three-body bound state with energy
\begin{equation}\label{E3universal}
E_3\simeq-4\hbar^2/ma_{p}^2.
\end{equation}
When $V_0$ is slightly less than $V_{c1}$, we indeed find that a three-body bound state appears. We have numerically solved the Schr\"{o}dinger equation to find the energies of the two-body and the three-body bound states with Gaussian pairwise interactions.
These energies are plotted in Fig.~\ref{fig:3bodybound}. We find that when $V_0$ is less than but close to $V_{c1}$,
these bound state energies are indeed close to the predictions of the aforementioned universal formulas.
%}

The one-dimensional square-barrier and square-well potentials and the Gaussian potential we have studied above are different
from the true interactions of ultracold atoms in quasi-one-dimensional (quasi-1D) optical wave guides in which the transverse motion of the atoms
is frozen to a length scale $a_\perp$ that is usually much larger than the characteristic range of the van der Waals potential
between the atoms. The 1D effective range $r_p$ of ultracold atoms in quasi-1D is much larger than the range of atomic interaction \cite{PhysRevLett.100.170404}, but the model potentials we have studied above have $r_p\sim r_0$. 
In Ref.~\cite{PhysRevA.93.023629} it is shown that the large 1D effective range has important consequences for the three-body states,
and in particular the ratio between the energies of the three-body shallow bound state and the two-body shallow bound state
deviates significantly from four at large and positive $a_p$ \cite{PhysRevA.93.023629}, in contrast to Eqs.~\eqref{E2universal} and \eqref{E3universal} in our paper.
Therefore the effect of the large 1D effective range on the three-body scattering hypervolume $D_F$
may also be large for real ultracold atoms.
The numerical calculation of $D_F$ for real ultracold atoms is expected to be much more difficult than the numerical
calculations in this paper: one would need to solve the three-body Schr\"{o}dinger equation in three dimensions.
We leave this as an open question.

\section{energy shifts due to $D_F$}\label{energy}
We consider three identical spin-polarized fermions on a line with length $L$, and impose the periodic boundary condition on the wave function: $\Psi(x_1+L,x_2,x_3)=\Psi(x_1,x_2,x_3)$. Consider an energy eigenstate in which the momenta of the fermions are $k_1$, $k_2$ and $k_3$ in the absence of interactions.
When we introduce interactions that give rise to a nonzero $D_F$, the shift of the energy eigenvalue due to a nonzero $D_F$ is
\begin{equation}\label{energy-3fermion}
\mathcal{E}_{k_1k_2k_3}=\frac{\hbar^2D_F}{12m L^2}(k_1-k_2)^2(k_2-k_3)^2(k_3-k_1)^2.
\end{equation}
See Appendix.~\ref{sec:energy} for the details of the derivation of this formula.

In addition, if there are two-body interactions, in general the shift of the energy of the three fermions will also contain terms due to the two-body parameters including $a_p,r_p$, etc.; nevertheless, the shift due to $D_F$ in \Eq{energy-3fermion} is still valid. We can also calculate the leading-order shift of the three-body energy due to $a_p$ by using a method similar to the one used in Appendix.~\ref{sec:energy}:
\begin{equation}
    \mathcal{E}_{k_1k_2k_3}^\text{2-body}=\frac{\hbar^2a_p}{mL}[(k_1-k_2)^2+(k_2-k_3)^2+(k_3-k_1)^2].
\end{equation}

We then generalize the energy shift to $N$ fermions in the periodic length $L$. The number density of the fermions is $n=N/L$.
We define the Fermi wave number $k_F=\pi n$, the Fermi energy $\epsilon_F=\hbar^2 k_F^2/2m$, and the Fermi temperature $T_F=\epsilon_F/k_B$, where $k_B$ is the Boltzmann constant. 

\subsection{Adiabatic shifts of energy and pressure in the thermodynamic limit due to $D_F$}
Starting from a many-body state at a finite temperature $T$, if we introduce a nonzero $D_F$ \emph{adiabatically}, the energy shift at first
order in $D_F$ is equal to the sum of the contributions from all the triples of fermions, namely
\begin{equation}
	\Delta E=\frac{1}{6}\sum_{k_1k_2k_3}\mathcal{E}_{k_1k_2k_3}\, n_{k_1}n_{k_2}n_{k_3},
\end{equation}
where $n_{k}=(e^{\beta(\epsilon_{k}-\mu)}+1)^{-1}$ is the Fermi-Dirac distribution function, $\beta=1/k_B T$, $\epsilon_{k}=\hbar^2 k^2/2m$ is the kinetic energy of a fermion, and $\mu$ is the chemical potential. The summation over $k$ can be replaced by a continuous integral $\sum_{k}=L \int dk/(2\pi)$ in the thermodynamic limit. Carrying out the integral, we get
\begin{align}
	\Delta E(T)&=\frac{N\hbar^2 D_F}{768\sqrt{\pi} m}k_F^8\nonumber\\
	 &\times\widetilde{T}^{9/2}\left[ 3 f_{1/2}(z)f_{3/2}(z)f_{5/2}(z)-f_{3/2}^3(z)\right] ,
\end{align}
where $\widetilde{T}=T/T_F$, $z=e^{\beta \mu}$, and the function $f_{\nu}(z)$ is defined as
\begin{equation}
	f_{\nu}(z)\equiv -\mathrm{Li}_{\nu}(-z)=\frac{2}{\Gamma(\nu)}\int_0^{\infty}\!\! dx~ \frac{x^{2\nu-1}}{1+e^{x^2}/z},
	%=z-\frac{z^2}{2^\nu}+\frac{z^3}{3^\nu}-\frac{z^4}{4^\nu}+\cdots
\end{equation}
where $\mathrm{Li}_{\nu}$ is the polylogarithm function.
The number of fermions satisfies
$
	N=\sum_{k} \frac{1}{e^{\beta (\epsilon_{k}-\mu)}+1},
$
and this leads to the equation of the chemical potential $\mu$:
\begin{equation}
\frac{2}{\sqrt{\pi}}=\sqrt{\widetilde{T}} ~f_{1/2}(\\e^{\widetilde{\mu}/\widetilde{T}}),
\end{equation}
where $\widetilde{\mu}=\mu/\epsilon_F$.

\begin{figure*}[htb]
\centering   	
\subfloat[Energy]
{   \label{ep1}
	\includegraphics[width=0.5\textwidth]{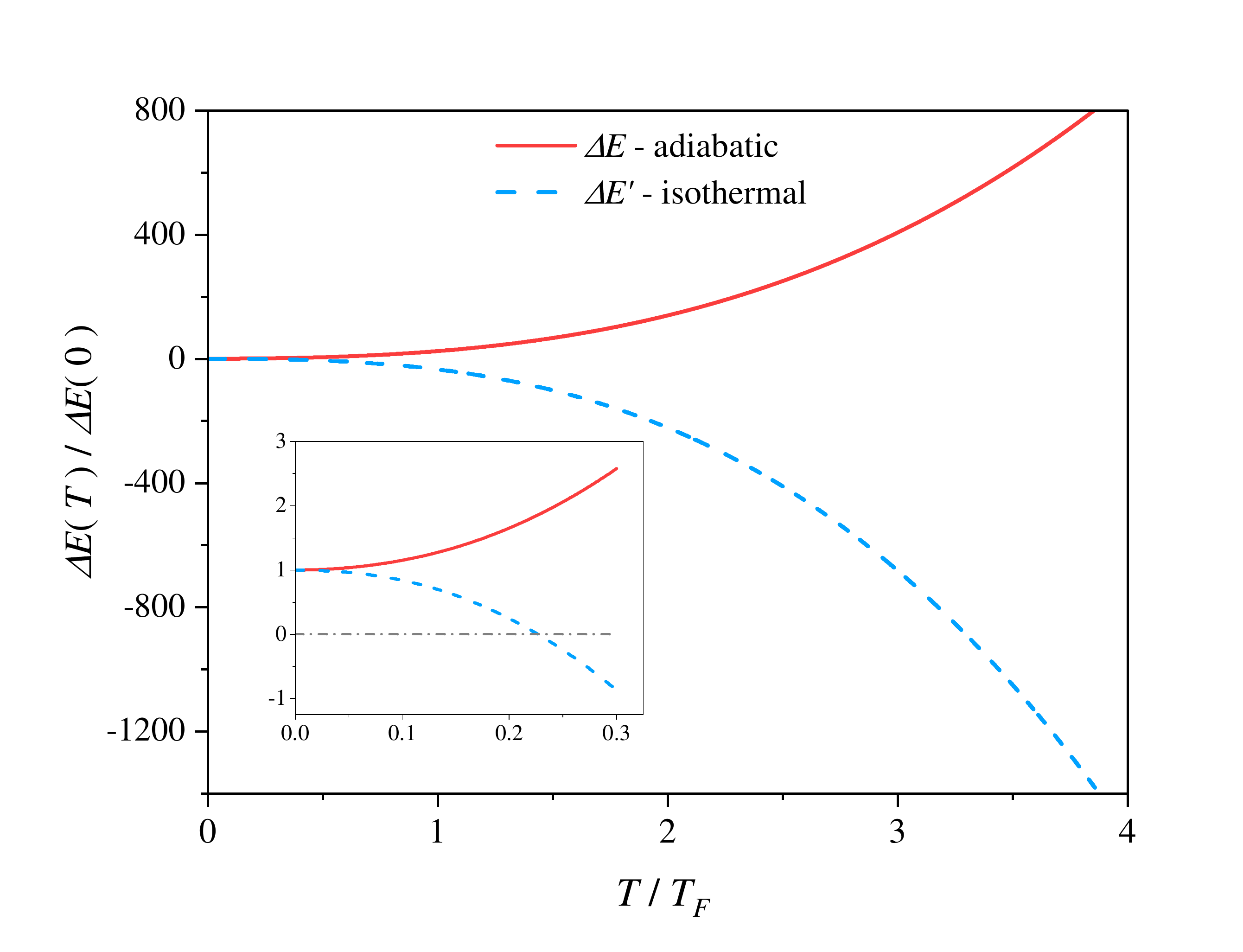}   
}
\subfloat[Pressure]
{   \label{ep2}
	\includegraphics[width=0.5\textwidth]{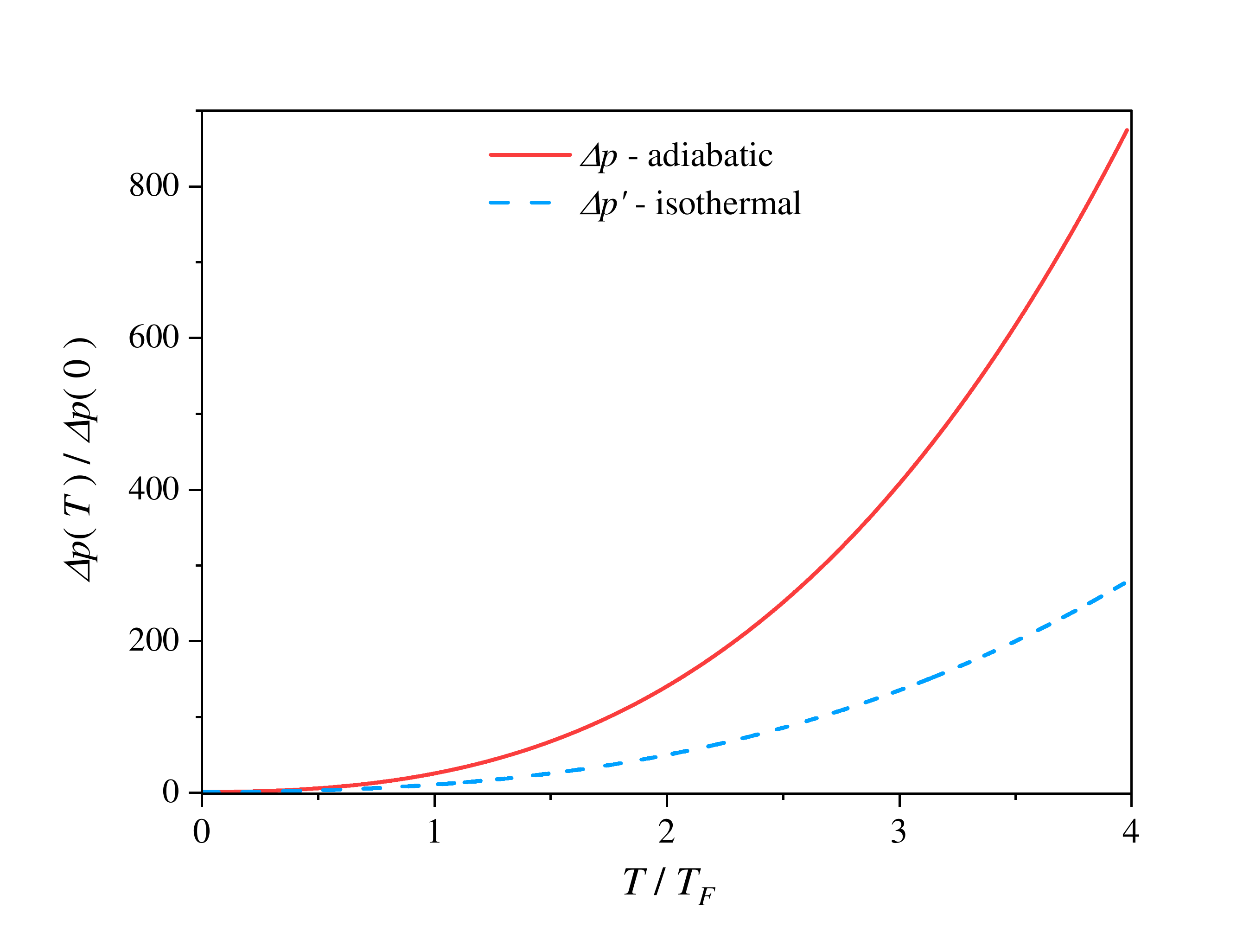} 
}	
\caption{The shifts of energy (a) and pressure (b) caused by the adiabatic (red solid lines) or isothermal (blue dashed lines) introduction of $D_F$ vs the temperature $T$. At $T\simeq0.2268T_F$, the isothermal energy shift $\Delta E$ changes sign.}
\label{fig:energy and pressure}
\end{figure*}

In the low temperature limit, namely $T\ll T_F$, 
\begin{equation}
	\Delta E(T)=\frac{N\hbar^2 D_F}{405\pi^2 m}k_F^8\left[ 1+\frac{3}{2}\pi^2 \widetilde{T}^2+%\frac{29\pi^4}{80}\widetilde{T}^4+
	\O(\widetilde{T}^4)\right].
\end{equation}
In an intermediate temperature regime, $T_F\ll T\ll T_e$, 
\begin{equation}
	\Delta E(T)=\frac{N\hbar^2 D_F}{48\pi^2 m}k_F^{8}\widetilde{T}^3
	\left[1+\frac{9}{4 \sqrt{2 \pi \widetilde{T}}}+\O(\widetilde{T}^{-1})\right],
\end{equation}
where $T_e=\frac{\hbar^2}{2mr_e^2k_B}$.
If $T$ is comparable to or higher than $T_e$, the de Broglie wave lengths of the fermions will be comparable to or shorter than the range $r_e$ of interparticle interaction potentials, and we can no longer use the effective parameter $D_F$ to describe the system.
See Fig. \ref{ep1} for $\Delta E$ as a function of the initial temperature.

The pressure of the spin-polarized Fermi gas changes by the following amount due to the adiabatic introduction of $D_F$:
\begin{equation}
		\Delta p=-\left(\frac{\partial \Delta E}{\partial L}\right)_{S,N}=\frac{8\Delta E}{L},\label{pressure adia}
\end{equation}
where the subscripts $S,N$ prescribe that we keep the entropy $S$ and the particle number $N$ fixed when taking the partial derivative. See Fig. \ref{ep2} for $\Delta p$ as a function of the initial temperature.

\subsection{Isothermal shifts of energy and pressure in the thermodynamic limit due to $D_F$}
If the interaction is introduced adiabaticly, the temperature will increase (if $D_F>0$) or decrease (if $D_F<0$). The change of temperature is
\begin{equation}
	\Delta T=\left(\frac{\partial \Delta E}{\partial S}\right)_{N,L}.
\end{equation}
So if we introduce $D_F$ isothermally, the energy shift $\Delta E'$ should be
\begin{equation}
	\Delta E'=\Delta E-C \Delta T=\left(1-T \frac{\partial}{\partial T}\right)\Delta E,
\end{equation}
where $C$ is the heat capacity of the noninteracting Fermi gas at constant volume.
In the low temperature limit, $T\ll T_F$,
\begin{equation}\label{DeltaE'lowT}
	\Delta E'(T)=\frac{N\hbar^2 D_F k_F^8}{405\pi^2 m}\left[ 1-\frac{3}{2}\pi^2 \widetilde{T}^2+%\frac{29\pi^4}{80}\widetilde{T}^4+
	\O(\widetilde{T}^4)\right].
\end{equation}
In an intermediate temperature regime, $T_F\ll T\ll T_e$, 
\begin{equation}\label{DeltaE'highT}
	\Delta E'(T)=\frac{N\hbar^2 D_F}{48\pi^2 m}k_F^{8}\widetilde{T}^3
	\left[-2-\frac{27}{8 \sqrt{2 \pi \widetilde{T}}}+\O(\widetilde{T}^{-1})\right].
\end{equation}
According to Eqs.~\eqref{DeltaE'lowT} and \eqref{DeltaE'highT}, $\Delta E'$ changes sign as we increase the temperature.
Therefore, there is a critical temperature $T_c$ at which $\Delta E'=0$. We find 
\begin{equation}
	T_c \simeq 0.2268 T_F.
\end{equation}

The pressure of the spin-polarized Fermi gas changes by the following amount due to the isothermal introduction of $D_F$:
\begin{equation}
	\Delta p'=\Delta p-\frac{2 C\Delta T}{L}=
	\left(1-\frac{1}{4}T \frac{\partial}{\partial T}\right)\Delta p.
\end{equation}
In the low temperature limit, $T\ll T_F$,
\begin{equation}
	\Delta p'=\frac{8n\hbar^2 D_F}{405\pi^2 m}k_F^{8}
	\left[1+\frac{3}{4}\pi^2 \widetilde{T}^2+O(\widetilde{T}^4)\right].
\end{equation}
In an intermediate temperature regime, $T_F\ll T\ll T_e$,
\begin{equation}
	\Delta p'=\frac{n\hbar^2 D_F}{6\pi^2 m}k_F^{8}\widetilde{T}^3
	\left[\frac{1}{4}+\frac{27}{32\sqrt{2\pi\widetilde{T}}}+O(\widetilde{T}^{-1})\right].
\end{equation}

The shifts of energy and pressure are plotted as functions of temperature in Fig.~\ref{ep1} and Fig.~\ref{ep2} respectively.

\section{The Three-body recombination rate\label{sec:recombination}}
If the collision of the three particles is purely elastic, $D_F$ is a real number.
But if the two-body interaction supports bound states, then the three-body collisions are usually not purely elastic, and the three-body recombination may occur. In this case $D_F$ becomes complex, and the three-body recombination rate constant is proportional to the imaginary part of $D_F$ \cite{zhu2017threebody,braaten2006universality}.

Within a short time $\Delta t$, the probability that no recombination occurs is $\mathrm{exp}(-2|\mathrm{Im} {E}|\Delta t/\hbar)\simeq 1-2|\mathrm{Im} {E}|\Delta t/\hbar$. Then the probability for one recombination is $2|\mathrm{Im} {E}|\Delta t/\hbar$.
Since each recombination event causes the loss of three low-energy fermions, the change of the number of remaining low-energy fermions in the short time $dt$ is
\begin{equation}
	d N=-\frac{1}{6}\sum_{k_1 k_2 k_3}3\frac{2d t}{\hbar}|\mathrm{Im}\mathcal{E}_{k_1k_2k_3}| n_{k_1}n_{k_2}n_{k_3}.
\end{equation}
This leads to
\begin{equation}
	\frac{d n}{d t}=-L_3 n^{3},
\end{equation}
and the coefficient $L_3$ is
\begin{align}
	L_3&=\frac{\pi^{3/2}}{128}\frac{\hbar |\mathrm{Im}D_F|}{m}k_F^6\nonumber\\
	&\times\widetilde{T}^{9/2}\left[ 3 f_{1/2}(z)f_{3/2}(z)f_{5/2}(z)-f_{3/2}^3(z)\right].
\end{align}
% \begin{widetext}
% \begin{align}
% 	L_3&=\frac{\pi^{3/2}}{128}\frac{\hbar |\mathrm{Im}D_F|}{m}k_F^6\times\widetilde{T}^{9/2}\left[ 3 f_{1/2}(z)f_{3/2}(z)f_{5/2}(z)-f_{3/2}^3(z)\right].
% \end{align}
% \end{widetext}
$L_3$ depends on the density $n$ and the temperature $T$.

In the low temperature limit, $T\ll T_F$,
\begin{equation}
	L_3\simeq \frac{2}{135}\left(1+\frac{3\pi^2}{2}\widetilde{T}^2\right)\frac{\hbar|\mathrm{Im}D_F|}{m}k_F^6.
\end{equation}
In particular, at $T=0$,
\begin{equation}
	L_3=\frac{2\hbar|\mathrm{Im} D_F|}{135m} k_F^6,
\end{equation}
and $L_3$ is proportional to $n^6$.

In an intermediate temperature regime, $T_F\ll T\ll T_e$, we find that
\begin{equation}\label{L3highT}
	L_3\simeq \frac{m^2}{\hbar^5}|\mathrm{Im}D_F|(k_B T)^3,
\end{equation}
and $L_3$ is approximately proportional to $T^3$, which is consistent with the prediction in Ref. \cite{Esry2007}.

\section{Summary and Discussion}
We derived the asymptotic expansions of the three-body wave function $\Psi$ for identical spin-polarized fermions colliding at zero energy in one dimension, and defined the three-body scattering hypervolume $D_F$. Now the scattering hypervolumes of spin-polarized fermions have been defined in 3D \cite{wang2021fermion3D}, 2D \cite{wang2022fermion2D} and 1D.
For weak interaction potentials, we derived an approximate formula for $D_F$ by using the Born expansion. 
For stronger interactions, one can solve the three-body Schr\"{o}dinger equation numerically at zero energy and match the resultant wave function with the asymptotic expansion formulas we have derived in this paper to numerically compute the values of $D_F$.
We did such numerical calculations for the square-barrier, square-well, and Gaussian potentials.

We considered three fermions along a line with periodic boundary condition
and derived the shifts of their energy eigenvalues due to a nonzero $D_F$, and then considered the dilute spin-polarized Fermi gas in 1D and derived the shifts of its energy and pressure due to a nonzero $D_F$.

Finally, we studied the dilute spin-polarized atomic Fermi gas in 1D with interaction potentials that support two-body bound states, for which we have three-body recombination processes and $D_F$ has nonzero imaginary part,
and we derived formulas for the three-body recombination rate constant $L_3$ in terms of the imaginary part of $D_F$ and the temperature and density of the Fermi gas.

One can similarly define the three-body scattering hypervolumes for identical bosons or for distinguishable particles in 1D and study their physical implications.

For ultracold atoms, one can use the optical lattice to confine them in quasi-1D, and the van der Waals range of the interatomic potential
is usually much shorter than the radial confinement length. One can solve the three-body problem in three dimensional space to numerically
determine the one-dimensional scattering hypervolume of the three atoms.

\begin{acknowledgments}
This work was supported by the National Key R\&D Program of China (Grants No.~2019YFA0308403 and No.~2021YFA1400902).
%(Grants No.~2021YFA1400902).
\end{acknowledgments}

\appendix
\section{Born Series for weak interactions\label{sec:Born}}
For weak-interaction potentials, we can expand the wave function as a Born series:
\begin{equation}
\Psi=\Psi_0+\widehat{G}\mathcal{V} \Psi_0+(\widehat{G}\mathcal{V})^2 \Psi_0 +\cdots,
\end{equation}
where $\Psi_0=s_1s_2s_3=s^3/4-sR^2$ is the wave function of three free fermions, $\widehat{G}=-\widehat{H}_0^{-1}$ is the Green's operator, $\widehat{H}_0$ is the three-body kinetic-energy operator. $\mathcal{V}=U(s_1,s_2,s_3)+\sum_i V(s_i)$, where $V(s_i)$ and $U$ are two-body and three-body finite-range potentials whose characteristic range is $r_e$.

\subsection{The first-order term}
The first order term $\Psi_1$ in the Born series is
\begin{align}
	\Psi_1=&\widehat{G}\mathcal{V}\Psi_0=\widehat{G}U\Psi_0+\sum_i \widehat{G}V_i\Psi_0\nonumber\\
	=&\frac{m}{\hbar^2}\int d^2 \bm{\xi}' \mathcal{G}(\bm{\xi}-\bm{\xi}')U(\bm{\xi}')\Psi_0(\bm{\xi}')\nonumber\\
	&+\frac{m}{\hbar^2}\sum_i \int d^2 \bm{\xi}' \mathcal{G}(\bm{\xi}-\bm{\xi}')V_i(\bm{\xi}')\Psi_0(\bm{\xi}'),
\end{align}
where $\bm{\xi}=(s,2R/\sqrt{3})$ and $\bm{\xi}'=(s',2R'/\sqrt{3})$ are two-dimensional vectors, and $V_i(\bm{\xi}')=V(s_i')$. Without losing generality, we set $s_2'=s'$, $s_1'=-\frac{1}{2}s'+R'$, $s_3'=-\frac{1}{2}s'-R'$.
$\mathcal{G}$ is the Green's function \cite{zhu2017threebody,wang2022fermion2D} in two-dimensional space which satisfies $\nabla_{\xi}^2 \mathcal{G}(\bm{\xi}-\bm{\xi}')=\delta(\bm{\xi}-\bm{\xi}')$, and its expression is
\begin{equation}
	\mathcal{G}(\bm{\xi}-\bm{\xi}')=\frac{1}{2\pi}\ln |\bm{\xi}-\bm{\xi}'|.
\end{equation}

We define
\begin{equation}
\Psi_{1i}\equiv\widehat{G}V_i\Psi_0.    
\end{equation}
For $i=2$, we have $V_i(\bm{\xi}')=V(s')$ and
\begin{align}
	\Psi_{1i}&=\frac{m}{\hbar^2}\int d^2 \bm{\xi}' ~\mathcal{G}(\bm{\xi}-\bm{\xi}')V_i(\bm{\xi}')\Psi_0(\bm{\xi}')\nonumber\\
	=&\frac{m}{\hbar^2}\int_{-\infty}^{\infty}\!\!dx'\int_{-\infty}^{\infty}\!\!dy'~ \frac{1}{16\pi}\ln\left[ (x_i-x')^2+(y_i-y')^2\right]\nonumber \\
	&\times V(x')(x'^3-3x' y'^2).\label{psi1i}
\end{align}
Here and in the following, we define $x_i=s_i$, $x'=s'$, $y_i=2R_i/\sqrt{3}$ and $y'=2R'/\sqrt{3}$ for simplicity.
We first integrate over $y'$ in \Eq{psi1i}. To avoid the divergence in the integral, we integrate over $y'$ from $-\lambda$ to $\lambda$ firstly and then take the limit $\lambda\rightarrow \infty$. 
Then we integrate over $x'$ and take the sum over $i$ to get
\begin{align}\label{sumpsi1i}
	\sum_i \Psi_{1i}=&\sum_i\left[ -\frac{1}{2}\alpha_3(x_i)+\frac{3}{4}\alpha_1(x_i)\left(y_i^2-x_i^2\right)\right. \nonumber\\
	&\left. -x_i\bar{\alpha}_2(x_i) -\frac{1}{4}\left(x_i^3-3x_i y_i^2\right)\bar{\alpha}_0(x_i)\right],
\end{align}
%We omit the terms which will be canceled after taking integral over $x'$ or after taking sum over $i$. We get
% \begin{equation}
% 	\Psi_{1i}=\frac{1}{16\pi}\int_{-\infty}^{+\infty}\!\!dx'~ V(x')\left[
% 	2\pi x'^3 |x_i-x'|+2\pi x' (|x_i-x'|^3-3y_i^2 |x_i-x'|)\right].
% \end{equation}
% Then we get
% \begin{equation}
% 	\Psi_{1i}=-\frac{1}{2}\alpha_3(x_i)+\frac{3}{4}\alpha_1(x_i)\left(y_i^2-x_i^2\right)-\bar{\alpha}_2(x_i) x_i-\frac{1}{4}\left(x_i^3-3x_i y_i^2\right)\bar{\alpha}_0(x_i).
% \end{equation}
where the functions $\alpha_n(x)$ and $\bar{\alpha}_n(x)$ at $x>0$ are defined as
\begin{subequations}
\begin{align}
	&\alpha_n(x)=\frac{m}{\hbar^2}\int_0^{x}\!dx' \:x'^{n+1}V(x'),\\
	&\bar{\alpha}_n(x)=\frac{m}{\hbar^2}\int_x^{\infty}\! dx' x'^{n+1} V(x').
\end{align}
\end{subequations}
At $x>r_e$, $\alpha_n(x)$ becomes a constant $\alpha_n$ and $\bar{\alpha}_n(x)=0$ because the potential $V(x')$ vanishes at $x'>r_e$. 
We also require $\alpha_n(x)$ to be odd functions and $\bar{\alpha}_n(x)$ to be even functions of $x$, namely
\begin{equation}
\alpha_n(-x)=-\alpha_n(x)	,~~~\bar{\alpha}_n(-x)=\bar{\alpha}_n(x).
\end{equation}
If $|x_1|$, $|x_2|$, and $|x_3|$ are all greater than $r_e$, \Eq{sumpsi1i} is simplified as
\begin{equation}
	\sum_i \Psi_{1i}=\sum_i\left[ -\frac{1}{2}\alpha_3+\frac{3}{4}\alpha_1\left(y_i^2-x_i^2\right)\right]\sgn(x_i).
\end{equation}

For any values of $x_i$,
\begin{align}\label{3bodyU}
    \widehat{G}U\Psi_0=&\frac{m}{\hbar^2}\int_{-\infty}^{\infty}\!\!\!dx'\int_{-\infty}^{\infty}\!\!\!dy'~\frac{1}{16\pi}\ln\left[ (x_i-x')^2+(y_i-y')^2\right] \nonumber\\
    &\times U(x',y')(x'^3-3x' y'^2).
\end{align}
Since $U$ is a finite-range potential, the integral on the right-hand side of \Eq{3bodyU} may be expanded when $x_i$ and $y_i$ go to infinity simultaneously. Expanding this integral at large $B$, we get
\begin{align}
    &\widehat{G}U\Psi_0\nonumber\\
    &\simeq \frac{m}{\hbar^2}\frac{-(x^3-3x y^2)}{24\pi(x^2+y^2)^3}\int_{-\infty}^{\infty}\!\!\!dx'\int_{-\infty}^{\infty}\!\!\!dy' ~U(x',y')(x'^3-3x' y'^2)^2\nonumber\\
    &=-\frac{m}{\hbar^2}\frac{3\sqrt{3}s_1s_2s_3}{4\pi B^6}\int_{-\infty}^{\infty}\!\!\!ds'\int_{-\infty}^{\infty}\!\!\!dR'~U(s',R') (s_1' s_2' s_3')^2\nonumber\\
    &\equiv-\frac{3\sqrt{3}s_1s_2s_3}{4\pi B^6}\Lambda.
\end{align}

\subsection{The second-order term}
The second-order term $\Psi_2$ in the Born series is
\begin{align}
\Psi_2 &= \widehat{G}\mathcal{V}\Psi_1 \nonumber\\
&= \sum_{ij}\widehat{G}V_i\widehat{G}V_j \Psi_0+\sum_i \widehat{G}V_i\widehat{G}U\Psi_0 +\sum_i \widehat{G}U\widehat{G}V_i\Psi_0\nonumber\\
&~~~~+(\widehat{G}U)^2\Psi_0,
\end{align}
We define 
\begin{align}
    &\Psi_{2,ij}=\widehat{G}V_i\widehat{G}V_j \Psi_0\nonumber\\
    &=\frac{m}{\hbar^2}\int\!\!\!\!\int\!dx'dy' \frac{1}{4\pi}\ln\left[ (x_i-x')^2+(y_i-y')^2\right]\nonumber\\
    &~~~~\times V(x')\Psi_{1j}(x',y').
\end{align}

In particular, if $j=i$,
\begin{align}
\Psi_{2,ii}=&\int\!\!\!\!\int\!dx'dy' \frac{1}{4\pi}\ln\left[ (x_i-x')^2+(y_i-y')^2\right] V(x')\nonumber\\
	&\times\left[-\frac{1}{2}\alpha_3(x')+\frac{3}{4}\alpha_1(x')\left(y'^2-x'^2\right)\right.\nonumber\\
	&\left.-\bar{\alpha}_2(x') x'-\frac{1}{4}\left(x'^3-3x' y'^2\right)\bar{\alpha}_0(x')\right].
\end{align}
If $|x_1|$, $|x_2|$, and $|x_3|$ are all greater than $r_e$, we can evaluate the integral to obtain
\begin{equation}
\Psi_{2,ii}=\left[ \beta_3-\frac{3}{4}\left(y_i^2-x_i^2\right)\beta_1\right] \sgn(x_i),
\end{equation}
where $\beta_1$ and $\beta_3$ are defined as
\begin{subequations}
	\begin{align}
		&\beta_1=\int_0^{\infty}\!\!\!dx\int_0^{x}\!\!\!dx'\: 2x x'^2 V(x)V(x'),\\
		&\beta_3=\int_0^{\infty}\!\!\!dx\int_0^{x}\!\!\!dx'\: (xx'^4+2x^3 x'^2 ) V(x)V(x').
	\end{align}
\end{subequations}

If $j\neq i$,
\begin{widetext}
\begin{align}
&\sum_{j\neq i}\Psi_{2,ij}=
\frac{1}{4\pi}\int\!\!\!\!\int\!dx'dy' \ln\left[ (x_i-x')^2+(y_i-y')^2\right] V(x')\nonumber\\
&\times \Big[
-\frac{1}{2}\alpha_3 (x_2')+\frac{3}{4}\alpha_3 (x_2')(y_2'^2-x_2'^2)-x_2'\bar{\alpha}_2(x_2')-\frac{1}{4}(x_2'^3-3x_2' y_2'^2)\bar{\alpha}_0(x_2')\nonumber\\
&~~~~-\frac{1}{2}\alpha_3 (x_3')+\frac{3}{4}\alpha_3 (x_3')(y_3'^2-x_3'^2)-x_2'\bar{\alpha}_2(x_3')-\frac{1}{4}(x_3'^3-3x_3' y_3'^2)\bar{\alpha}_0(x_3')\Big],
\end{align}
where $x_2'=-\frac{1}{2}x'+\frac{\sqrt{3}}{2}y',~~y_2'=-\frac{\sqrt{3}}{2}x'-\frac{1}{2}y',~~x_3'=-\frac{1}{2}x'-\frac{\sqrt{3}}{2}y',~~y_3'=+\frac{\sqrt{3}}{2}x'-\frac{1}{2}y'.$

\begin{align}
	\sum_{j\neq i}\Psi_{2,ij}=&
	\frac{1}{4\pi}\int\!\!\!\!\int\!dx'dy' V(x')\left[ \frac{1}{2}\alpha_3 (\frac{\sqrt{3}}{2}y')-\alpha_1 (\frac{\sqrt{3}}{2}y')(x'^2-\frac{3}{8}y'^2)+\frac{\sqrt{3}}{2}y'\bar{\alpha}_2 (\frac{\sqrt{3}}{2}y')-\frac{\sqrt{3}}{2}x'^2 y' \bar{\alpha}_0 (\frac{\sqrt{3}}{2}y')\right] \nonumber\\
	&\times\left\lbrace 
	\ln\left[ (x_i-x')^2+\big(y_i-y'+\frac{1}{\sqrt{3}}x'\big)^2\right]-\ln\left[ (x_i-x')^2+\big(y_i-y'-\frac{1}{\sqrt{3}}x'\big)^2\right]\right\rbrace \nonumber\\	
	+&\frac{1}{4\pi}\int\!\!\!\!\int\!dx'dy' V(x')\left[ \frac{\sqrt{3}}{2}x'y'\alpha_1 (\frac{\sqrt{3}}{2}y')+\frac{3}{4}x' y'^2 \bar{\alpha}_0(\frac{\sqrt{3}}{2}y')\right] \nonumber\\
	&\times \left\lbrace 
	\ln\left[ (x_i-x')^2+\big(y_i-y'+\frac{1}{\sqrt{3}}x'\big)^2\right]+\ln\left[ (x_i-x')^2+\big(y_i-y'-\frac{1}{\sqrt{3}}x'\big)^2\right]\right\rbrace. 
\end{align}

We define
\begin{align}
&\frac{1}{2}\alpha_3 (\frac{\sqrt{3}}{2}y')-\alpha_1 (\frac{\sqrt{3}}{2}y')(x'^2-\frac{3}{8}y'^2)+\frac{\sqrt{3}}{2}y'\bar{\alpha}_2 (\frac{\sqrt{3}}{2}y')-\frac{\sqrt{3}}{2}x'^2 y' \bar{\alpha}_0 (\frac{\sqrt{3}}{2}y')\nonumber\\
\equiv&\frac{1}{2}\alpha_3~\sgn(y')-\alpha_1(x'^2-\frac{3}{8}y'^2)~\sgn(y')+f_1 (x',y'),
\end{align}
where
\begin{equation}
f_1(x',y')=f_{11}(y')+x'^2 f_{12}(y'),
\end{equation}
and
\begin{align}
&f_{11}(y')=\frac{1}{2}\alpha_3 (\frac{\sqrt{3}}{2}y')-\frac{1}{2}\alpha_3~\sgn(y')+\frac{3}{8}y'^2\alpha_1(\frac{\sqrt{3}}{2}y')-\frac{3}{8} y'^2\alpha_1~\sgn(y')+\frac{\sqrt{3}}{2}y' \bar{\alpha}_2(\frac{\sqrt{3}}{2}y'),\\
&f_{12}(y')=-\alpha_1(\frac{\sqrt{3}}{2}y')+\alpha_1~\sgn(y')-\frac{\sqrt{3}}{2}y' \bar{\alpha}_0(\frac{\sqrt{3}}{2}y').
\end{align}
$f_{11}$ and $f_{12}$ are odd functions of $y'$. They are short-range functions, namely they vanish at $\sqrt{3}|y'|/2>r_e$.

We also define
\begin{equation}
 \frac{\sqrt{3}}{2}x'y'\alpha_1 (\frac{\sqrt{3}}{2}y')+\frac{3}{4}x' y'^2 \bar{\alpha}_0(\frac{\sqrt{3}}{2}y')
 \equiv
 \frac{\sqrt{3}}{2}\alpha_1 x' y' ~\sgn(y')+x' f_2 (y').
\end{equation}
\begin{equation}
f_2(y')=\frac{\sqrt{3}}{2}y'\alpha_1 (\frac{\sqrt{3}}{2}y')+\frac{3}{4} y'^2 \bar{\alpha}_0(\frac{\sqrt{3}}{2}y')
-
\frac{\sqrt{3}}{2}  y'\alpha_1,
\end{equation}
$f_{2}$ is an even function of $y'$. It vanishes at $\sqrt{3}|y'|/2>r_e$.
Then
\begin{align}
	\sum_{j\neq i}\Psi_{2,ij}=&
	\frac{1}{4\pi}\int\!\!\!\!\int\!dx'dy' V(x')\left[ \frac{1}{2}\alpha_3~\sgn(y')-\alpha_1(x'^2-\frac{3}{8}y'^2)~\sgn(y')+f_{11}(y')+x'^2 f_{12}(y')\right] \nonumber\\
	&\times\left\lbrace 
	\ln\left[ (x_i-x')^2+\big(y_i-y'+\frac{1}{\sqrt{3}}x'\big)^2\right]-\ln\left[ (x_i-x')^2+\big(y_i-y'-\frac{1}{\sqrt{3}}x'\big)^2\right]\right\rbrace \nonumber\\	
	+&\frac{1}{4\pi}\int\!\!\!\!\int\!dx'dy' V(x')\left[\frac{\sqrt{3}}{2}\alpha_1 x' y' ~\sgn(y')+x' f_2 (y')\right] \nonumber\\
	&\times \left\lbrace 
	\ln\left[ (x_i-x')^2+\big(y_i-y'+\frac{1}{\sqrt{3}}x'\big)^2\right]+\ln\left[ (x_i-x')^2+\big(y_i-y'-\frac{1}{\sqrt{3}}x'\big)^2\right]\right\rbrace .
\end{align}

For large $x_i$ and $y_i$, we get
\begin{align}
	\sum_{j\neq i}\Psi_{2,ij} =& -\frac{3\sqrt{3}}{\pi}\alpha_1^2 y_i \theta_i~\sgn(x_i)
	-\frac{(x_i^3-3x_i y_i^2)}{\sqrt{3}\pi (x_i^2+y_i^2)^3}\left( \frac{20}{9}\alpha_3^2-\frac{28}{45}\alpha_1 \alpha_5\right)\nonumber\\
	& +(\text{terms which will be canceled after summation over }i)+O(B^{-4}).
\end{align}
We have not evaluated the terms $\widehat{G}V_i \widehat{G}U\Psi_0, ~\widehat{G}U \widehat{G}V_i\Psi_0$ and $(\widehat{G}U)^2\Psi_0$ in the Born series. The full expression of $\Psi_2$ is 
\begin{align}
\Psi_2=&\sum_i\left[  \beta_3-\frac{3}{4}\left(y_i^2-x_i^2\right)\beta_1
-\frac{3\sqrt{3}}{\pi}\alpha_1^2 y_i \theta_i\right] \sgn(x_i)
-\frac{(x^3-3x y^2)}{\sqrt{3}\pi (x^2+y^2)^3}\left( \frac{20}{3}\alpha_3^2-\frac{28}{15}\alpha_1 \alpha_5\right)+O(V^2B^{-9})\\
&+O(UV)+O(U^2),\nonumber
\end{align}
as is shown in the main text.
\end{widetext}
Comparing the resultant Born series and the 111 expansion of the wave function, we get
\begin{subequations}
\begin{align}
    a_p &=\alpha_1-\beta_1+O(V^3).\\
    a_p^2 r_p &=-\frac{2}{3}\alpha_3+\frac{4}{3}\beta_3+O(V^3).
\end{align}
\end{subequations}
and
\begin{align}
    D_F &=\frac{\Lambda}{2}+\frac{5}{2}\alpha_3^2-\frac{7}{10}\alpha_1 \alpha_5+O(V^3)\nonumber\\
    &~~~+O(UV)+O(U^2).
\end{align}
For the square well potential with strength $V_0$ and range $r_e=1$,
\begin{subequations}
\begin{align}
&\alpha_1=\frac{1}{3}V_0,~~\alpha_3=\frac{1}{5}V_0,~~\alpha_5=\frac{1}{7}V_0,\\
&\beta_1=\frac{2}{15}V_0^2,~~\beta_3=\frac{13}{105}V_0^2.
\end{align}
\end{subequations}
Substituting these results, we get
\begin{subequations}
\begin{align}
    a_p&=\frac{1}{3}V_0-\frac{2}{15}V_0^2+O(V_0^3),\\
    a_p^2 r_p&=-\frac{2}{15}V_0+\frac{52}{315}V_0^2+O(V_0^3),
\end{align}
\end{subequations}
which are consistent with the direct calculation of $a_p$ and $r_p$.
For the scattering hypervolume, we have
\begin{equation}
D_F=\frac{1}{15}V_0^2+O(V_0^3),
\end{equation}
which is consistent with the result of numerical computations for small $V_0$.

\section{Shifts of the energy of three fermions in one dimension with periodic boundary conditions due to $D_F$}\label{sec:energy}

The normalized wave function of three free fermions with momenta $\hbar k_1,\hbar k_2,\hbar k_3$ in a large periodic line with length $L$ is
\begin{equation}\label{freePsi}
	\Psi_{k_1 k_2 k_3}=\frac{1}{\sqrt{6}L^{3/2}}
	\left| \begin{matrix}
		e^{i k_1 x_1}& e^{i k_1 x_2} & e^{i k_1 x_3}\\
		e^{i k_2 x_1}& e^{i k_2 x_2} &e^{i k_2 x_3}  \\ 
		e^{i k_3 x_1}& e^{i k_3 x_2} & e^{i k_3 x_3}
	\end{matrix}\right|  .
\end{equation}
We define the Jacobi momenta $\hbar q,\hbar p,\hbar k_c$ such that
\begin{subequations}
	\begin{align}
		&k_1=\frac{1}{3}k_c+\frac{1}{2}q+p,\\
		&k_2=\frac{1}{3}k_c+\frac{1}{2}q-p,\\
		&k_3=\frac{1}{3}k_c-q.
	\end{align}
\end{subequations}
$\hbar k_c$ is the total momentum of three fermions. We extract the motion of the center of mass $R_c=(x_1+x_2+x_3)/3$, 
\begin{equation}
	\Psi_{k_1 k_2 k_3}=
	\frac{1}{\sqrt{L}}e^{i k_c\cdot R_c}\Phi_{p,q}.\label{C3}
\end{equation}
Suppose that the typical momentum of each fermion is $\sim2\pi\hbar/\lambda$.
For small hyperradii, $B\ll\lambda$, we Taylor expand $\Phi_{p,q}$ and get
\begin{equation}
	\Phi_{p,q}\simeq \frac{-i}{\sqrt{6}L}\left( p^3-\frac{9}{4}p q^2\right)\left( \frac{1}{4}s^3-s R^2\right) .\label{Phiqp}
\end{equation}
$\Phi_{p,q}$ is the wave function of the relative motion of three free fermions. If we introduce a small three-body $D_F$ adiabatically, $\Phi_{p,q}$ is changed to
\begin{align}\label{relativePhi}
	\Phi_{p,q}\simeq \frac{-i}{\sqrt{6}L}\Big( p^3-\frac{9}{4}p q^2\Big)\Big( \frac{1}{4}s^3-s R^2\Big)\bigg( 1-\frac{3\sqrt{3}D_F}{2\pi B^6}\bigg) 
\end{align}
for $r_e\ll B\ll\lambda$. The wave function satisfies the free Schr\"{o}dinger equation outside of the range of interaction,
\begin{equation}\label{SEinbox}
	-\frac{\hbar^2}{m}\nabla_{\bm\xi}^2\Phi_{p,q}=E\Phi_{p,q},
\end{equation}
where $\bm\xi=(s,2R/\sqrt{3})$ is a two-dimensional vector, $E$ is the energy of the relative motion, and $B=\sqrt{3}\xi/2$.

For large values of $L$, we may compute the energy $E$ approximately. We rewrite \Eq{SEinbox} as
\begin{subequations}
	\begin{align}
		&-\frac{\hbar^2}{m}\nabla_{\bm\xi}^2\Phi_{1}=E_1\Phi_{1},\label{e1}\\
		&-\frac{\hbar^2}{m}\nabla_{\bm\xi}^2\Phi_{2}^{*}=E_2\Phi_{2}^{*},\label{e2}
	\end{align}
\end{subequations}
for two slightly different interactions that yield two slightly different scattering hypervolumes, $D_{F1}$ and $D_{F2}$ respectively.
Here we omit the subscript $p,q$ for simplicity. Multiplying both sides of \Eq{e1} by $\Phi_2^*$,
multiplying both sides of \Eq{e2} by $\Phi_1$, subtracting the two resultant equations, and taking two-dimensional integral over $\bm\xi$ for $\xi>\xi_0$
(where $\xi_0$ is any length scale satisfying $r_e\ll\xi_0\ll\lambda$), we get
%\begin{subequations}
\begin{align}
	&-\frac{\hbar^2}{m}\int_{\xi>\xi_0}\!\!\! d^2 \xi~ \bm{\nabla}_{\bm\xi}\cdot(\Phi_2^*\bm{\nabla}_{\bm\xi} \Phi_1-\Phi_1\bm{\nabla}_{\bm\xi}\Phi_2^*)\nonumber\\
	&=(E_1-E_2)\int_{\xi>\xi_0}\!\!\! d^2\xi~ \Phi_1 \Phi_2^*.\label{Gauss}
\end{align}
%\end{subequations}
In the bulk part of the configuration space, $\Phi_{1}\simeq\Phi_2$. 
Note also that the wave function for the relative motion is normalized, and that
the volume of the region $\xi<\xi_0$ is small and may be omitted in the normalization integral.
So the right hand side of \Eq{Gauss} is
\begin{equation}
	\frac{2}{\sqrt{3}}(E_1-E_2)\int_{\xi>\xi_0}\!\!\! ds ~dR~ |\Phi|^2
	\simeq\frac{2}{\sqrt{3}}(E_1-E_2).
\end{equation}
Applying Gauss's theorem to the left hand side of \Eq{Gauss}, we get
\begin{equation}\label{intonS}
	-\frac{\hbar^2}{m}\oint_{\xi=\xi_0}\!\!\! d\vect{S}\cdot (\Phi_2^*\bm{\nabla}_{\bm\xi} \Phi_1-\Phi_1\bm{\nabla}_{\bm\xi}\Phi_2^*)\simeq\frac{2}{\sqrt{3}}(E_1-E_2),
\end{equation}
where $S$ is the surface of the circle with radius $\xi=\xi_0$ centered at the origin,
and $d\vect S$ points toward the center of the circle.

To evaluate the integral on the circle $\xi=\xi_0$, we parametrize $\bm\xi=(\xi^{(1)},\xi^{(2)})$ as
\begin{subequations}
	\begin{align}
		&\xi^{(1)}=\xi \cos \varphi,\\
		&\xi^{(2)}=\xi \sin\varphi,
	\end{align}
\end{subequations}
where $0\leqslant\varphi<2\pi$. Here
$\xi^{(1)}=s$ and $\xi^{(2)}=2R/\sqrt3$.
%$R=\sqrt{3}\xi^{(2)}/2$. 
The surface element $d\vect{S}$ is
\begin{align}
	d\vect{S}=&- \bm{\xi} d\varphi.
\end{align}
The minus sign in the above equation means that the direction of $d\vect{S}$ is towards the origin.  
Assuming that $\Phi_1$ and $\Phi_2$ satisfy \Eq{relativePhi} with $D_F=D_{F1}$ and $D_F=D_{F2}$ respectively,
and evaluating the integral in \Eq{intonS} on the circle with radius $\xi=\xi_0$, we get
\begin{align}
	&E_1-E_2=\frac{\hbar^2}{3m L^2}(D_{F1}-D_{F2})\left(p^3-\frac{9}{4}p q^2\right)^2\nonumber\\
	&=\frac{\hbar^2}{12m L^2}(D_{F1}-D_{F2})(k_1-k_2)^2(k_2-k_3)^2(k_3-k_1)^2.
\end{align}
This result agrees with \Eq{energy-3fermion} in the main text.

\bibliography{ref}
\end{document}